%% file: manuscript.tex
\documentclass[aps,prb,twocolumn,letterpaper,superscriptaddress,showpacs,amsmath,citeautoscript]{revtex4-1}
\usepackage{shellesc}
\usepackage[slantedGreek]{mathptmx}
\usepackage[scaled]{helvet}
\usepackage{mathcomp}
\usepackage{upgreek}
\usepackage[colorlinks,citecolor=blue,filecolor=blue,linkcolor=blue,urlcolor=blue]{hyperref}

\usepackage{tabularx,booktabs}
\AtBeginDocument{
  \heavyrulewidth=.08em
  \lightrulewidth=.05em
  \cmidrulewidth=.03em
  \belowrulesep=.65ex
  \belowbottomsep=0pt
  \aboverulesep=.4ex
  \abovetopsep=0pt
  \cmidrulesep=\doublerulesep
  \cmidrulekern=.5em
  \defaultaddspace=.5em
}

\usepackage{tikz}
\usepackage{pgfplots}
\usetikzlibrary{backgrounds}
\usetikzlibrary{arrows}
\usepgfplotslibrary{groupplots}
\pgfplotsset{compat=newest}
\usetikzlibrary{spy}
\newlength\figureheight
\newlength\figurewidth

\usetikzlibrary{patterns}

\newlength\imagewidth
\newlength\imagescale

\newcommand{\upe}{\text{e}}

\newcommand*{\microscope}{Dresden \emph{in-situ} (S)TEM special}

\begin{document}

\title{The \microscope\ with a continuous-flow liquid-helium cryostat}

\author{Felix B\"{o}rrnert}
\email{f.boerrnert@ifw-dresden.de}

\affiliation{Leibniz-Institut f\"{u}r Festk\"{o}rper- und Werkstoffforschung  Dresden e.\,V., Helmholtzstra\ss e 20, 01069 Dresden, Germany}

\author{Felix Kern}
\affiliation{Leibniz-Institut f\"{u}r Festk\"{o}rper- und Werkstoffforschung  Dresden e.\,V., Helmholtzstra\ss e 20, 01069 Dresden, Germany}

\author{Franziska Seifert}
\affiliation{Leibniz-Institut f\"{u}r Festk\"{o}rper- und Werkstoffforschung  Dresden e.\,V., Helmholtzstra\ss e 20, 01069 Dresden, Germany}%

\author{Thomas Riedel}
\affiliation{CEOS GmbH, Englerstra\ss e 28, 69128 Heidelberg, Germany}

\author{Heiko M\"{u}ller}
\affiliation{CEOS GmbH, Englerstra\ss e 28, 69128 Heidelberg, Germany}

\author{Bernd B\"{u}chner}
\affiliation{Leibniz-Institut f\"{u}r Festk\"{o}rper- und Werkstoffforschung  Dresden e.\,V., Helmholtzstra\ss e 20, 01069 Dresden, Germany}

\author{Axel Lubk}
\affiliation{Leibniz-Institut f\"{u}r Festk\"{o}rper- und Werkstoffforschung  Dresden e.\,V., Helmholtzstra\ss e 20, 01069 Dresden, Germany}

\begin{abstract}
  Fundamental solid state physics phenomena typically occur at very low temperatures, requiring liquid helium cooling in experimental studies. Transmission electron microscopy is a well-established characterization method, which allows probing crucial materials properties down to nanometer and even atomic resolution. Due to the limited space in the object plane, however, suitable liquid-helium cooling is very challenging. To overcome this limitation, resolving power was sacrificed in our \microscope, resulting in more than 60\:mm usable experimental space in all directions with the specimen in the center. With the installation of a continuous-flow liquid-helium cryostat, any temperature between 6.5\:K and 400\:K can be set precisely and kept for days. The information limit of the \microscope\ is about 5\:nm. It is shown that the resolution of the \microscope\ is currently not limited by aberrations, but by external instabilities, that are currently addressed.
\end{abstract}

\maketitle

\section{Introduction}

Transmission electron microscopy (TEM) in its various flavours including scanning TEM (STEM) and electron holo\-graphy is a well-established characterization technique, which allows to resolve the local structure, chemical composition, as well as static and dynamic electromagnetic fields within solids down to atomic length scales.\cite{Haider1998,Reimer2008,Bosman2007,Egerton2009,Frankel1970,Lichte1991,Linck2012,Winkler2018} Consequently, (S)TEM is ubiquitous in materials science, chemistry, and life sciences.

For its application in condensed matter physics, however, liquid helium (lHe) cooling of the sample inside the transmission electron microscope is often mandatory because fundamental solid-state effects, such as superconductivity, quantum hall effect, charge density waves, or frustrated magnetism are suppressed by thermal fluctuations at elevated temperatures.\cite{Ashcroft2011} Moreover, secondary radiation damage effects originating from radiolysis are reduced when cooling the sample, providing a strong impetus for the implementation of sample cooling into TEM of biological materials \cite{Egerton2012,Chiu1986}. Mainly because of the stabilisation of shock-frozen biological samples, TEM at low sample temperatures has been actively developed for a couple of decades. One milestone in that development was the introduction of superconducting magnetic coils by the cooling of the whole objective lens including the pole pieces surrounding the specimen.\cite{Laberrigue1964,Boersch1966} Due to the enormous technological and development efforts including large running costs, superconducting objective lenses were eventually abandoned in favour of more flexible and cheaper lHe sample holders, only cooling at the very tip of the holder containing the specimen. However, the small distance between the objective lens pole pieces severely hampers the installation of a vibration-free and long-time stable lHe cryostat. Practically, current designs allow cooling times in the order of one hour only and generally suffer from drift and mechanical stability issues. \cite{Gatan2018}

Here, we report on the \microscope, where we replaced a whole section of the original microscope column\,---\,formerly including the objective lens pole pieces and the sample stage\,---\,by a large experimental chamber. Nanometer spatial resolution is maintained by employing (S)TEM hardware aberration correction. Into the experimental chamber, we fitted a continuous-flow lHe cryostat, through which any temperature between 6.5\:K and 400\:K can be set precisely and kept for days. We discuss the performance of the microscope including the cryostat.

\section{Outline of the \microscope}
\label{sec:Setup-of-the}

The general idea behind the \microscope\ is to provide all capabilities of (S)TEM methods to a comparably large experimental space for complex \textit{in-situ} experiments compromising spatial resolving power.\cite{Boerrnert2015,Boerrnert2016} For convenience, the experimental space should be accessible from several large ports that bear standard vacuum flanges.

The first plug-in to the \microscope\ is a cryostat that fulfills the demands for solid-state physics experiments, namely, a base temperature below 10\:K, precise temperature control in a wide temperature range, temperature stability over several hours, drift and vibration levels below image resolution, no drift induced by temperature change, at least four cooled electrical feedthroughs for transport measurements, and optional incorporation of at least two mobile electrical probers. Of course, the cryostat must not disturb the operation of the microscope; one of the challenges here is that the cold sample has to be movable in all spatial directions with sufficiently small step resolution and additionally, one needs a small sample tilting capability to be able to avoid dynamic contrast in electron holographic imaging modes.

\begin{figure}
  \includegraphics[width=\columnwidth]{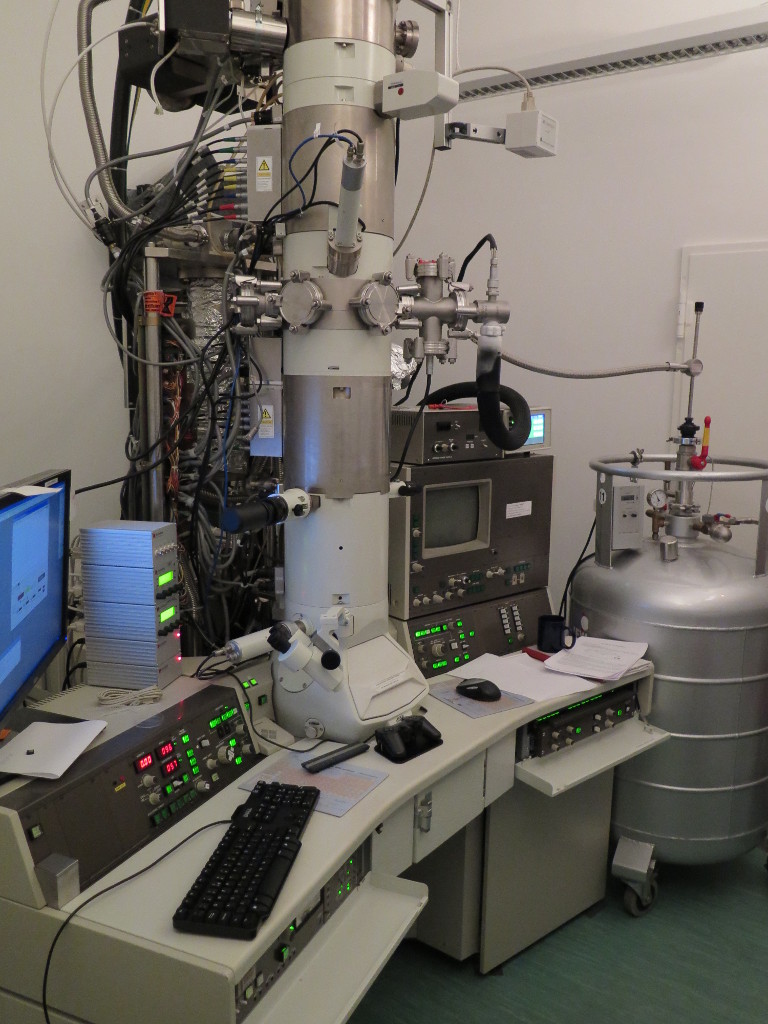}
  \caption{Photograph of the \microscope\ with the cryostat in operation.}
  \label{fig.mic}
\end{figure}

The basis for the \microscope\ is a JEOL\:JEM-2010F retrofitted with CEOS probe and image-side aberration correctors, a fiber-coupled CCD camera, and a digital scan generator.\cite{Boerrnert2013} Additional non-standard attachments include a secondary electron detector and a M\"{o}llenstedt-D\"{u}ker bi-prism. The fully operational \microscope\ is shown in figure\:\ref{fig.mic}.

\subsection{Optical setup}
\label{subsec:Polepieces-of-the}

In terms of electron optics, the most pervasive modification to the \microscope\ is the effective removal of the objective lens \citep{Boerrnert2015}, thereby expanding the space around the specimen for installing various instrumentation.

For TEM imaging, we use the first transfer doublet of the image-side hexapole-type aberration corrector as the main image-forming lens. As we do not alter the geometry of the setup, the working distance increases from 1\:mm to about 60\:mm, enlarging the chromatic and geometric aberrations equivalently, with the latter ones largely compensated by the corrector. The attainable resolution and magnification in this so-called ``pseudo-Lorentz'' mode \cite{Snoeck2006,Houdellier2008} are decreased by approximately one order of magnitude (see ref.\:\onlinecite{Boerrnert2015} for a detailed assessment of the optical performance immediately after the removal of the pole pieces), which is suitable for medium resolution imaging.

\begin{figure}
  \input{figures/cetcor/cetcor}
  \caption{Schematic ray paths for the standard (\emph{dashed lines}) and pseudo-Lorentz (\emph{solid lines}) operation mode of the hexapole corrector. The transfer lenses (TL) and the main correction hexapoles (HP) are indicated on the right side, the other adjustment elements on the left. The underlying figure is a reproduction from reference \onlinecite{Haider2009}, where also the remaining abbreviations are explained.}
  \label{fig:CETCOR}
\end{figure}
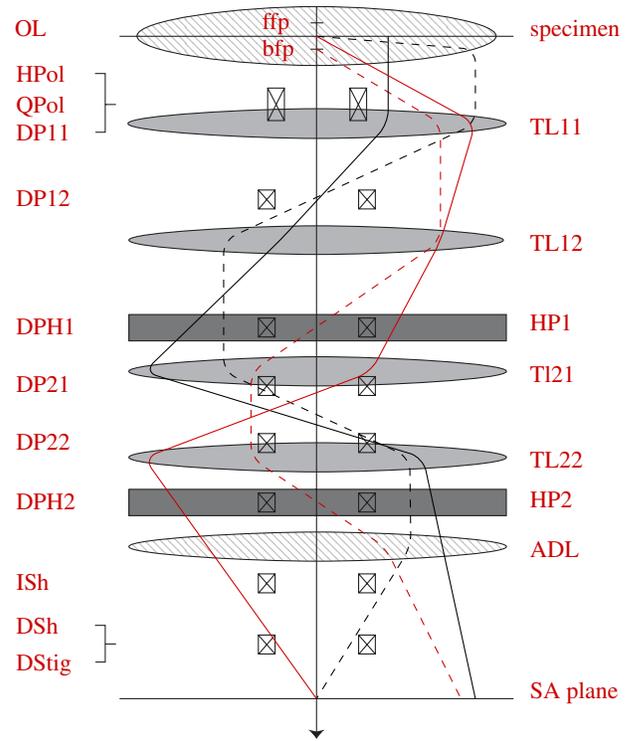

Of course, after the objective lens removal all imaging modes have to be reconfigured because the optical planes behind the sample are shifted with respect to the conventional modes. Especially, the corrector is not operated in standard mode, because of the use of the first telescopic transfer lens doublet as objective lens substitute. Here, the back focal plane of the new objective lens does not coincide with the first and second hexapole plane, and the ray paths are not (anti)symmetric with respect to the mid-plane of the corrector (as in the standard mode, see figure\:\ref{fig:CETCOR}) \cite{Mueller2008}.

A practical consequence of this new alignment is, that the defocus is now controlled with the help of TL12  lens in the corrector. Also, the third-order spherical aberration is changed by directly accessing the main correction elements HP1 and HP2 \emph{via} changing their excitation strength in parallel. Generally, besides the third-order spherical aberration, only the second order aberrations are corrected, since the higher order aberrations do not influence the imaging in this mode. The correction procedures for the first- and second-order aberrations are similar to the standard routines.

The aforementioned reduction of the magnification in positions space leads to an increased magnification in diffraction mode, \emph{i.\,e.}, an increase in camera length. This increase in camera length can be problematic for the recording of large diffraction patterns, \emph{e.\,g.}, to determine the orientation or symmetry of the sample. A remedy consists of using the corrector's adapter optics for further post-magnification of the image to decrease the size of the diffraction pattern.

The TEM ``parallel'' illumination, normally formed by the condenser system in conjunction with the objective lens pre-field, is now provided with the help of the condenser-mini lens. For a proper operation of the aberration corrector, a certain beam tilt capability is necessary, because the geometrical aberrations to be corrected have to be measured first by means of a Zemlin tilt-tableau. In our setup we achieve a maximum tilt angle of $10$\:mrad. Zemlin tableaus recorded for aberration determination in high-resolution (S)TEM, normally contain tilt angles of 20\:mrad and higher, however, due to the optical constraints in pseudo-Lorentz mode, 10 \:mrad is more than enough. We also tested an illumination setup, where the complete probe corrector including the condenser-mini lens was switched off and the last active optical element before the sample was the second condenser lens, residing approximately 0.4\:m above the sample plane. Here, the maximum tilt of $\sim 5$\:mrad was at the border of being usable for an aberration measurement. Note, however, that this is no fundamental optical limitation but only due to the use of an optical setup that was not designed for this operation.

In spite of focussing on conventional TEM here, we also note that a pseudo-Lorentz STEM mode has been aligned, making use of the reciprocity principle. Accordingly, the setup of the STEM corrector is very similar to that of the TEM corrector with the ray paths inverted. All TEM and STEM alignments were carried out for a primary electron acceleration voltage of $200$\:kV as well as $80$\:kV.

\subsection{The experiment chamber}
\label{subsec:Specimen-Chamber}

\begin{figure}
  \input{figures/chamber/chamber}
  \caption{Photograph of the new specimen chamber for the \microscope\ in a test setup, viewing from the top along the optical axis. All radial ports are fitted with standard ISO-K\:63 flanges with an inner diameter of $63$\:mm except the one in 6-o'clock direction that is the flange for the ion getter pump of the original microscope. The thin permalloy layer shields against external magnetic stray fields.}
  \label{fig:specimen-chamber}
\end{figure}
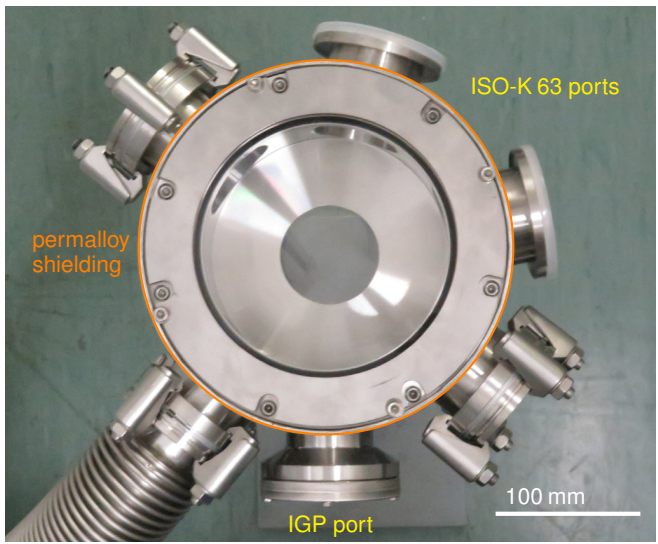

To fully exploit the experimental space gained from removing the objective lens' pole pieces, a new specimen chamber has been designed (\emph{cf.}\:figure\:\ref{fig:specimen-chamber}). It features five ports equipped with standard ISO-K\:63 vacuum flanges, to ensure that various instrumentation for \textit{in-situ} experiments can be adopted easily. Three of these ports are oriented in 90\textdegree\ angles towards each other (that is, two of them offer a straight line through the column crossing the electron beam path) and the remaining two in odd angles in order to maximise the number of ports and because of the connection geometry to the original microscope column. A dedicated sixth port is designed to fit to the ion getter pump of the original microscope. The specimen chamber provides usable space of 70\:mm along the optical axis due to the geometry of the original microscope. The space usable for experiments is effectively dictated by the full width of the access flanges that is $63$\:mm.

To compensate for the reduced magnetic shielding due to removal of soft magnetic pole pieces as well as the yoke of the original specimen chamber, the new chamber is surrounded by a $1.5$\:mm thick permalloy shield. The shield was designed to reach the maximum shielding possible without blocking the access ports. An estimated shielding factor on the optical axis of about 40 with respect to external fields penetrating the chamber parallel to the sample plane could be achieved. Note, however, that it is difficult to assess impact of the fields emerging from the interrupted yoke of the objective lens, which may also disturb the beam paths. Shielding against external magnetic fields is particularly important in the \microscope, because of the long working distance of the imaging lens. The penetrating stray fields lead to an image spread reducing the spatial resolution \cite{uhlemann2013}, which will be discussed in detail further below.

In the design phase of the specimen chamber, we decided against the incorporation of valves separating the microscope column vacuum from the sample chamber. The two main reasons were the space such valves would have required and the technical complexity of possible solutions, not mentioning the costs. Since the new specimen chamber has no airlock, each change of the specimen requires the whole microscope column to be vented and evacuated again. While the vacuum system of the original microscope would be capable of handling the evacuation, it would take inconveniently long times and would quickly wear the ion getter pumps. Therefore, a cascade of two serially connected turbo molecular pumps and a rotary vane pump are connected in parallel to the specimen chamber to speed up the evacuation process. This turbo molecular pump cascade is separated from the specimen chamber by a full-width gate valve that can be operated independently from the microscopes vacuum system. Depending on the installed holder and its previous treatment this pump cascade makes it possible to change the specimen in about 1\:hour. The whole second pump cascade including the gate valve is set up computer-controllable and thus, is fully automatable.

\subsection{The automatisation hub}
\label{subsec.automatisation}

Because in multi-stimuli \emph{in-situ} experiments many instances have to be controlled simultaneously, centralised automatisation of the whole setup is mandatory. We face a wide range of instruments to be controlled with a similarly wide range of connection types and protocols. Therefore, the electronics workshop of the IFW Dresden developed an ethernet-based system with modular hardware connection converters. The whole system is laid out to be able to monitor and steer arbitrary hardware attached in real time and record a wide range of data.

These measures also ensure that the data obtained can be correlated correctly afterwards.

\subsection{The stage}
\label{subsec:stage}

To enable the sample movement in transmission electron  microscopes, remote stages are used exclusively. Moreover, standard TEM stages are equipped with an air lock to reduce the specimen transfer times. Constructing a comparable stage that offers a clear diameter in the magnitude that is meaningful for the task would be a huge technological challenge by itself and result in a very bulky and expensive device. The basic demands for a stage include, for instance, a clear beam path in all positions, a step width in the order of the resolving power of the instrument, a total travel distance in the order of millimeters, travel in all three space directions, and at least a small sample tilting capability in order to prevent dynamic conditions.

Because of the estimated efforts and costs of a conventional remote stage development, we decided to use an internal stage with a very small footprint that can be transferred to any new experimental setup. Here, additional demands are operation in all orientations (hanging, vertical, \ldots), being non-magnetic and vacuum compatible.

Our solution consists of a customised version of the LT3310 substage manufactured by Kleindiek Nanotechnik GmbH \cite{Kleindiek_2017}. The basic stage offers lateral movement of 10\:mm in both directions with a nominal resolution of 0.25\:nm with a drift of 1\:nm/min. The mobile stage plate offers a square hole with 12.5\:mm side length that is completely free in vertical direction in every position of lateral movement. The motors driving the movement are piezo based linear motors with the rough movement using a slip--stick mechanism and the whole device is non-magnetic and high vacuum compatible. The load capability in any position and orientation is 25\:g.

\begin{figure}
  \includegraphics[width=\columnwidth]{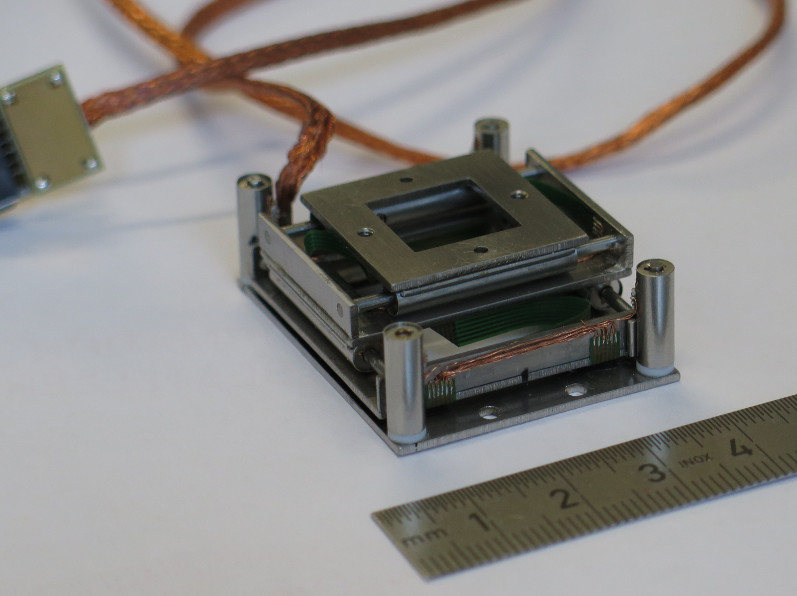}
  \caption{Photograph of the sample stage.}
  \label{fig:stage}
\end{figure}

As a custom design, the whole device is mounted on a base plate with a vertical linear piezo motor on each of the four corners offering 3\:mm total travel distance in the vertical dimension. Every motor is controllable individually. Also, the rooting of the motors is designed slightly flexible. Together, both design details allow to tilt the sample slightly in any direction. The base area of the whole stage is $33\times 40$\:mm$^2$, as can be seen in figure\:\ref{fig:stage}.

The stage offers as an extra feature four electrical feedthroughs, where the terminal contacts are moved with the stage.

\subsection{The cryostat}
\label{subsec:Kryo-Holder-and}

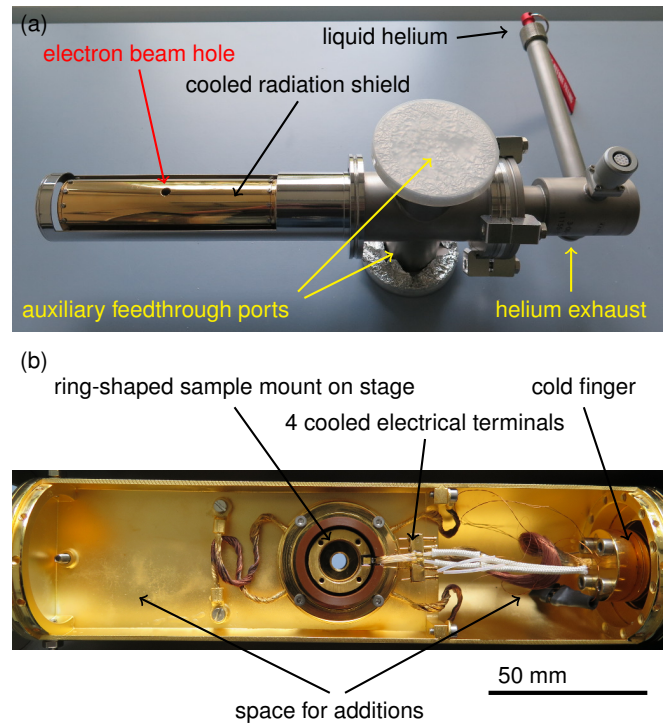
\begin{figure}
  \input{figures/cryo_layout/cryo_layout}
  \caption{The cryostat plug-in for the \microscope\ with the crucial parts indicated. (a) Photograph in the as-delivered state, (b) inner layout with both radiation shields unmounted.}
  \label{fig:Cryostat}
\end{figure}

The cryostat plug-in for the \microscope\ consists of four major parts: A standard, off-the-shelf continuous-flow cold-finger cryostat, model KONTI from CryoVac GmbH, a mechanical support frame, the portable sample stage described in the section above, and the stage-isolated movable cold sample mount. The complete plug-in is shown in figure\:\ref{fig:Cryostat}(a). 

The commercial cold-finger cryostat features an efficient heat exchanger providing constant lHe temperature with very low lHe consumption when cold. Continuous cooling below 10\:K for five days is possible if necessary using a full 100\:l lHe vessel. (If one needs longer cold standing time and does not require very low temperatures, the cryostat can be operated with liquid nitrogen as well.) Additionally, a second heat exchanger uses the exhaust He gas to cool the radiation shields to $\sim 20$\:K. The cryostat is steered by a proportional-integral-derivative controller, model CryoVac TIC\:500, with several temperature sensor inputs and heating outputs. With the controller, the system is not only automated but also completely computer-controllable.

The actual sample resides in a ring-shaped sample mount that is thermally connected to the cold finger by a copper braid, see figure\:\ref{fig:Cryostat}(b). Both, the ring shape and the braid connection are to prevent thermal-expansion-induced sample drift and also to decouple the sample from the vibrations of the heat exchanger. Attached to the copper braid socket on the sample mount side is a comb with four attachment points for electrical leads. Inside the ring-shaped sample mount, at the nearest feasible point to the sample, is a resistive heating and a temperature sensor. The sample mount resides in an isolating Vespel labyrinth-pot (the brown ring around the golden-coloured sample mount in figure\:\ref{fig:Cryostat}(b)) on a disk that is thermally connected to the exhaust-gas-cooled radiation shield and that can bear an optional second intermediate cooled radiation shield (not mounted in the figure). The whole circular construction is mounted on the stage described above isolated by another double-walled tube-shaped Vespel labyrinth, because the stage is always at room temperature. Therefore, the stage itself is also outside the large cooled radiation shield and the Vespel tube leads through a hole in this shield moving the sample mount inside.

The whole cryostat including the radiation shield is held by a massive steel frame with the  mount point far away from the sample position to dampen any vibrations induced by the helium flow. The stage resides on the frame carrying just the thermally isolated sample mount that is not touching the radiation shield. Additionally, the frame features two free standard ISO-K\:63 vacuum flanges that can be used, \emph{e.\,g.}, for auxiliary feedthroughs.

Conceptually, the cryostat has two operation mode configurations, a lowest-base-temperature mode with both radiation shields mounted and an easy-access mode with only a clamp-ring instead of the inner radiation shield. In the latter configuration, the sample resides on a flat cold surface that can be accessed from every direction of the top half-sphere. Also, the space in the frame is planned in a way that it can host two independent mobile electrical probers if one does not mount the larger radiation shield lid. Of course, the achievable base temperature will be compromised by thermal radiation in this configuration.

In full operation, the cryostat plug-in does introduce only little mechanical drift or vibration to the sample, judged in the order of the optical resolving power of the microscope, also not upon temperature change. The image resolution is the same in the warm state, all connectors removed, and in the cold state with helium flow. Also, the use of the resistive heating has no detectable effect on the imaging.

The base temperature at the sample mount achievable in the fully shielded configuration is 6.5\:K (20\:K in the easy access configuration). The maximum sample temperature that can be safely installed is 400\:K. In the low-temperature regime, the temperature stabilisation time after a temperature set-point change is less than one minute. The temperature stability is much better than 0.01\:K in the fully shielded configuration and better than 0.05\:K in the easy access configuration. 

Finally, the very local temperature of the sample at the observed region can significantly differ from the installed temperature due to local beam heating effects depending on the sample material and geometry.

\section{Electron optical performance}
\label{sec:Electron-Optical-Performance}

In order to characterize the optical performance of the TEM we took two separate complementary approaches. First, we measured the resolving power with the help of a suitable crystalline sample with a large lattice constant. Second, we measured all relevant aberrations, calculated the thus achievable information limit and compare it to the former result.

\subsection{Imaging resolution test}
\label{subsec:Young-Fringes-Test}

\begin{figure}
  \input{figures/cryo_resolution/cryo_resolution}
  \caption{Image of a catalase crystal taken at 200\:kV electron acceleration voltage. The images were taken at a specimen temperature of 6.5\:K.}
  \label{fig:catalase}
\end{figure}
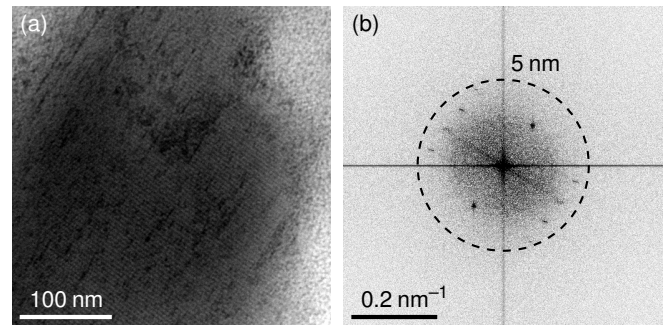

An estimate of the microscope's information limit can be obtained from imaging a suitable test sample. We use catalase because of its large lattice constants that yields signals in the image's Fourier transform at 0.114\:nm$^{-1}$ and 0.146\:nm$^{-1}$ for intrinsic calibration. An actual image taken at a sample temperature of 6.5\:K with 200\:kV electron acceleration voltage is shown in figure\:\ref{fig:catalase}. The signals in the Fourier transform extend to a spatial frequency of 0.2\:nm$^{-1}$, thus, the information limit of the \microscope\ with the cryostat plug-in at 200\:kV electron acceleration voltage is 5\:nm. The image resolution is isotropic. Also, the image acquisition time was set to 8\:s, in order to show that there is no drift. Images taken with the same configuration but at room temperature with all connections taken off the cryostat plug-in show the same imaging characteristics, excluding mechanical vibrations influencing the resolving power. The systematic reflections that occur in the Fourier transform can not stem from nonlinear contrast transfer because they are of first order, \emph{i.\,e.}, not a combination of lower-order reflections. Therefore, the information limit given here is not over-estimated due to nonlinear contrast transfer.

With a test frame holding just the sample stage without the cryostat, an image resolution better than 3\:nm could be obtained. Therefore, the cryostat plug-in has an influence on the imaging performance that is not due to mechanical vibrations.

\subsection{Lens aberrations and information limit}
\label{subsec:Determination-and-Correction}

The determination of the geometric aberrations in TEM mode was done with the standard Zemlin tableau method using the CEOS corrector software. Due to the general optical properties of the pseudo-Lorentz mode, a determination of the geometric aberrations only up to the second order and the third-order spherical aberration and astigmatism is possible. Using the manual correction, third-order spherical aberration values of a few tens of millimeters are achievable, compared to values in the range of ten meters in the uncorrected state.

The determination of the first order chromatic aberration coefficient $C_c$ is based on its definition \emph{via} the relation of the defocus change $\Delta C_1$ to a change of the electron acceleration voltage $\Delta U/U$:
\begin{equation}
\Delta C_1=C_{c}\frac{\Delta U}{U}.\label{eq:determination of the Cc}
\end{equation}
Here, the electron acceleration voltage was changed over a range of 1\:kV in 0.1\:kV steps and the induced defocus change was measured with the help of the Thon ring fit routine of the corrector software. A linear fit of the measurement results yielded the chromatic aberration coefficient. All aberration values measured are listed in table\:\ref{tab:aberrations}.

\begin{table}
\caption{Aberrations measured for the \microscope\ after hardware aberration correction.}
\label{tab:aberrations}
\begin{tabularx}{\columnwidth}{Xrr}
  \toprule
aberration type & at 200\:kV / mm  & \quad at 80\:kV / mm\\
\midrule
two-fold astigmatism & 0.0 (0.0) & 0.0 (0.0)\\
three-fold astigmatism & 0.0 (0.6) & 0.4 (0.3)\\
second-order axial coma & 0.2 (0.5) & 0.0 (0.2)\\
third-order spherical aberration & 46 (126) & 56 (105)\\
\midrule
chromatic aberration & 184 (2) & 129 (1) \\
\bottomrule
\end{tabularx}
\end{table}

Using the measured aberration values in combination with the electron source characteristics, the spatial coherence and the temporal coherence dampening envelope function can be calculated. Other dampening envelope functions might be generated by electro-magnetic stray fields and mechanical vibrations. The product of all dampening envelope functions yields the optical information limit and thus, the ultimate resolving power of the instrument \cite{Frank1973}.

The spatial coherence dampening envelope has the form
\begin{eqnarray}
E_{\text{s}} (q) &=& \upe ^{-\frac{1}{2}\left(\frac{2\uppi}{\lambda}\ \sigma_{\alpha}\ \nabla \chi\right)^2},\\   \nabla \chi &=& C_1 (\lambda q) + C_3 (\lambda q)^3 + C_5 (\lambda q)^5 + \ldots,  \nonumber
\end{eqnarray}
with $\lambda$ the electron wavelength, $\sigma_{\alpha}$ the illumination semi-angle distribution, $\chi$ the aberration function, $C_i$ the round geometric aberration coefficients ($C_1$ being the defocus), and $(\lambda q)$ the scattering angle. Here, we are taking in account only the round geometric aberrations $C_i$ up to the third order. Correspondingly, an approximation of the temporal coherence dampening envelope function has the form
\begin{eqnarray}
  E_{\text{t}} (q) &=& \upe ^{-\frac{1}{2}\left(\frac{2\uppi}{\lambda}\ \sigma_{\text{fs}}\ \frac{\partial \chi}{\partial C_1}\right)^2},\quad \tfrac{\partial \chi}{\partial C_1} = \tfrac{1}{2}(\lambda q)^2,\label{eq.fs_basic}
\end{eqnarray}
with the focus spread
\begin{equation}
  \sigma_{\text{fs}} = \sqrt{C_c^2 \left(\tfrac{\Delta U}{U}\right)^2 + C_c^2 \left(\tfrac{\Delta E_e}{eU}\right)^2 + \sigma_{\text{fs,residual}}^2}
  \label{eq.fs_old}
\end{equation}
with the total chromatic aberration coefficient $C_c$, the high tension stability $\left(\frac{\Delta U}{U}\right)$, the electron emission energy distribution $\Delta E_e$, the elementary charge $e$, and the objective lens current stability $\left(\tfrac{\Delta I}{I}\right)$. In our case, the residual focus spread $\sigma_{\text{fs,residual}}$, stemming from lens instabilities, can be safely neglected.

The information limit is the point where the total dampening envelope function, \emph{i.\,e.}, the product of all contributing envelope functions, enters the noise level. The noise level is conventionally assumed as $\upe^{-2}$ of the total contrast. This convention is still meaningful today because the noise is dominated by the Poisson noise of the imaging electrons themselves, and thus, the noise level is mainly dependent on the total electron dose in the image regardless of the detector. The optical information limit due to the spatial and chromatic dampening we determined to be 0.79\:nm$^{-1}$, corresponding to about 1.3\:nm resolving power.

Experimentally, we observed a significant reduction of image resolution down to 3\:nm with respect to this optical limit (using a test frame for the sample stage), which we ascribe to an additional image spread due to external magnetic fields.

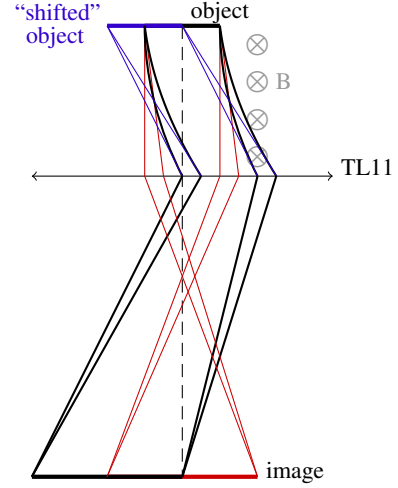
\begin{figure}
  \input{figures/imagespread/imagespread}
  \caption{Scheme for the origin of the image shift induced by magnetic fields between object and the imaging lens. Accordingly, a common parabolic deflection (\textit{black}) of the original beams (\textit{red}) due to a (correlated) magnetic field leads to a shifted image (indicated by \textit{blue} rays when referred back to the object plane), and hence an image spread, when integrated over random magnetic fields oscillations.}
  \label{fig:imagespread}
\end{figure}

To model the action of stray fields on the imaging process in lateral direction, we employ the image spread description\cite{uhlemann2013}. Accordingly, the thereby introduced resolution decrease is described by an envelope function in Fourier space
\begin{eqnarray}
E_{\text{is}} (q) &=& \upe ^{-\frac{1}{2}\left(\frac{2\uppi}{\lambda}\ \sigma_{\text{is}}\ (\lambda q)\right)^2}
\end{eqnarray}
with the image spread 
\begin{eqnarray}
\sigma_{\text{is}} &=& \frac{e\lambda}{2h} \zeta^2 \sigma_B ,
\end{eqnarray}
in position space with $h$ denoting Planck's constant, $\zeta$ the unshielded beam path, and $\sigma_B$ the magnetic stray field variation in lateral direction. Here, the factor 1/2 follows from back-tracking the bent trajectories to the object plane (indicated by blue lines in figure\:\ref{fig:imagespread}). We measured the magnetic field variation to be $15$\:nT and account for a dampening factor of 40 by the shields \cite{Boerrnert2015}, that is, we estimate a magnetic field variation of about 0.4\:nT at the optical axis. The working distance of the TL11 lens is about 60\:mm; this serves as a rough estimate of the unshielded path length. With these values, we determine an image spread of $0.43$\:nm.

\begin{figure}
  \input{figures/ctf/fs_dampening}
  \caption{Dampening envelope functions for the \microscope\ operated at 200\:kV electron acceleration voltage. \textit{green}\,---\,partial coherence dampening envelopes, with spatial coherence (\textit{dotted}) and temporal coherence (\textit{dashed}) envelope function. The parameters used for these calculations are $C_{c}=184$\:mm, $\left(\frac{\Delta U}{U}\right)=10^{-6}$, $\Delta E_e=0.34$\:eV (equivalent to $0.8$\:eV FWHM), $\sigma_{\alpha}=0.15$\:mrad, $C_3=46$\:mm and its corresponding Scherzer defocus. \textit{blue}\,---\,image spread dampening envelope function due to magnetic stray fields with $\zeta=60$\:mm and $\sigma_B=0.4$\:nT. \textit{gray}\,---\,noise level.}
  \label{fig:ctf}
\end{figure}
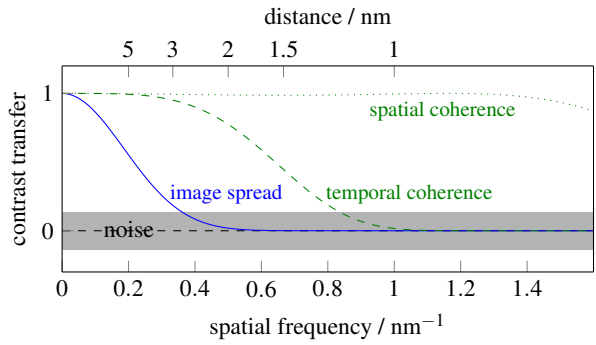

Figure\:\ref{fig:ctf} shows the different dampening envelope functions. As mentioned before, the optical information limit without disturbances is about 1.3\:nm. Due to the image spread we just estimated, the image resolution would be reduced to 1.64\:nm. Our measurements with a test frame bearing the sample stage result into a resolving power that is worse by a factor of approximately two. All in all, the estimate is a very rough one, since the unshielded path length is probably shorter. Also, the remaining magnetic yokes of the original objective lens might lead to magnetic disturbances preferentially to the beam path. Additionally, it might bend disturbances parallel to the optical axis into the lateral plane. Finally, the further reduction of the resolving power using the cryostat plug-in is most probably due to the cryostat playing the role of an antenna.

\section{Setup of Advanced Imaging Modes}
\label{sec:Setup-of-Advanced}

In addition to the reconfiguration of the standard imaging and diffraction modes, a dark-field mode, an off-axis holographic imaging mode, and different modes with varying excitation of the original objective lens coils were created. In the latter case, the excitation of the objective lens is employed in order to generate a magnetic field in the specimen plane to be used as external stimulus that is mainly oriented along the optical axis.

\subsection{Dark field imaging}
\label{subsec:Dark-Field-Mode}

\begin{figure}
  \input{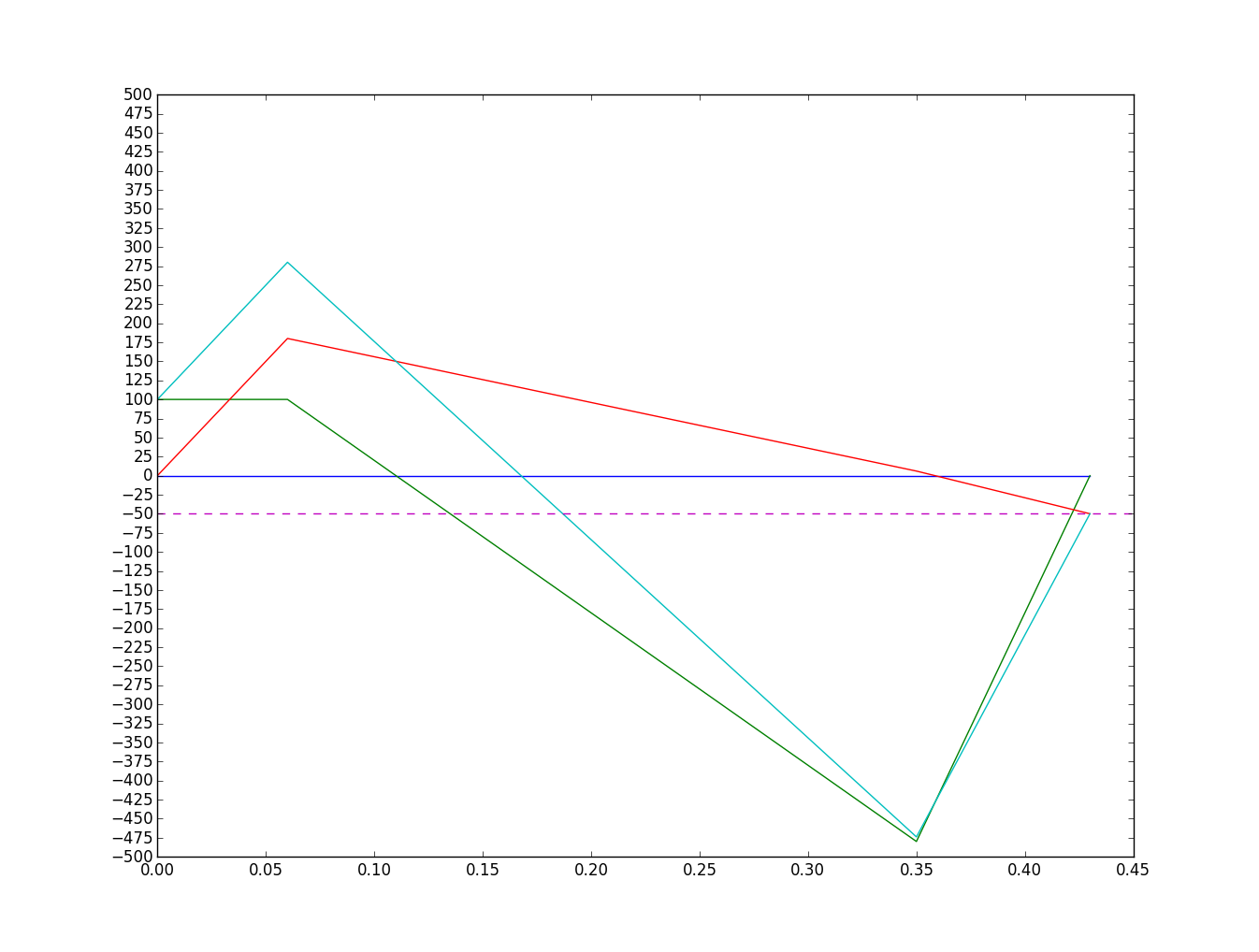}
  \caption{Calculated path of rays for one possible darkfield imaging mode employing the first transfer lens of the corrector and the final adaptive lens only (the rest is grayed out therefore). The horizontal axis is scaled such that the distance between the diffraction orders (emerging under an angle of 3\:mrad in the specimen plane) in the SA plane is 50\:\textmu m.}
  \label{fig:darkfield}
\end{figure}

For dark-field imaging or imaging with a contrast aperture, the original setup can not be used since the objective lens' back focal plane is shifted and the original objective aperture is gone. Nevertheless, dark field imaging is possible by reassigning the function of the objective aperture to the original selected-area-diffraction aperture. To that end the diffraction plane needs to be shifted to the selected-area (SA) plane, which can be accomplished by adapting the various transfer lenses of the Cs-corrector (see figure\:\ref{fig:CETCOR}). Note that these modes are not spherical aberration corrected, which is, however, not problematic because they are mainly employed in the medium resolution regime. As an example, we show a setting where the different diffraction orders of Au can be selected by a 50\:\textmu m aperture in the SA plane (given in figure\:\ref{fig:darkfield}). There is a multitude of corrector lens excitations facilitating the same splitting of diffraction orders in the SA plane; in practice, additional criteria such as a minimal ray pitch further narrow the range of suitable setups.

\subsection{Holographic Imaging Mode}
\label{subsec:Holographic-Imaging-Mode}

In the off-axis holographic imaging mode an electrostatic biprism is inserted in an intermediate image plane, where usually the selected-area-diffraction apertures are placed. This image plane needs to be shifted slightly, so that the biprism is located above or below the SA plane at a distance in the millimeter range. Otherwise, the two holographic partial waves are not overlapping in the image and can not interfere\cite{Lichte1996}. Usually, the biprism is placed above the SA plane and a positive voltage $U_{F}$ is applied. If it is placed below, a negative biprism voltage is necessary for the partial waves to overlap \cite{Sickmann2011}. 

In order to enhance the contrast $C=\left(I_{max}+I_{min}\right)\left(I_{max}-I_{min}\right)^{-1}$ of the interference fringes, an elliptical illumination was created using the condenser stigmators. The elliptical illumination increases the interference fringe contrast by an enhancement of the partial spatial coherence between the interfering partial waves in the direction of their overlap. Further information concerning the elliptical illumination is given in \citep{Lehmann2004}. The effective size of the illumination was optimised by a variation of the spot size and of the excitation of the second condenser lens. Employing these settings, we achieved hologram widths up to 900\:nm with fringe contrasts ranging from 10 $\%$ to 40 $\%$. That performance is comparable to other TEM instruments working in Lorentz mode. Larger fields of view can be generated by changing the biprism--plane distance and thereby compromising hologram resolution; hologram widths up to 5\:\textmu m have been tested.

\subsection{Magnetic field at sample position}
\label{subsec:Objective-Lens-Mode}

The removal of the objective lens from the optics and the replacement of the corresponding microscope column section was done in an economic way, that is, only such parts were removed that blocked the space needed. In the special case of the objective lens, the original coils are still in place. Exciting these coils after the removal of the pole pieces is not advisable, as it still has imaging properties but these are affected by strong (non-isoplanatic) aberrations due to the inappropriate shape of the magnetic field. Nonetheless, they may be excited in order to create a magnetic field in the specimen position as an external stimulus. For example, such fields may be used to stabilize a magnetic phase or to drive magnetic samples through hysteresis loops within \emph{in-situ} TEM experiments. To account for this additional imaging element, the excitation of the first transfer lens was reduced in order to maintain a focused image, while exciting the objective lens coils. Depending on the excitation degree of the objective lens coils, there are strong distortions visible, indicating the large magnitude of off-axial aberrations present in this mode.

The magnetic field, corresponding to the different excitation levels of the original objective lens coils was determined with the help of a teslameter, that was placed at approximately the sample position. Five new alignnments were created for the operation of the \microscope\ with excited original objective lens' coils. By raising the objective lens' coil excitation to the maximum value, two crossovers could be observed, which implies that the objective lens still acts similar to a $k^{2}=3$ lens \citep{Reimer2008}. Between those crossovers are stable regions, where imaging is possible in spite of large non-isoplanatic aberrations.

The resulting magnetic fields in the specimen plane parallel to the optical axis for the five imaging modes with excited objective lens' coils are 108\:mT, 123\:mT, 134\:mT, 160\:mT, and 204\:mT. Due to the electronics of the microscope that are used to excite the coils, the smallest magnetic field at sample position practically achievable by this method is 65\:mT, the correspondingly largest magnetic field is 204\:mT. If a lower or higher magnetic field is needed, it may be adjusted by using an external power controller in combination with the coils of the original objective lens or by a different magnetic field generating installation. 

\section{Temperature Calibration}
\label{sec:Applications-of-Cryo}

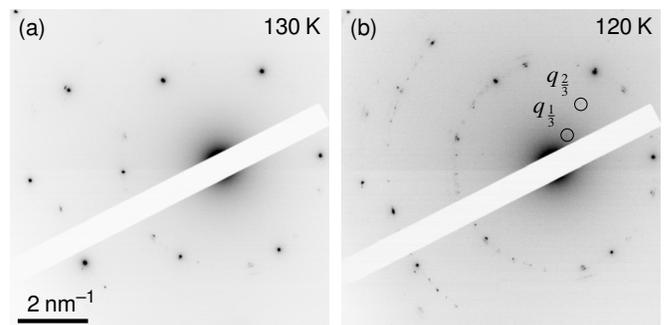
\begin{figure}
  \input{figures/calibtemp/calibtemp_2}
  \caption{Temperature series of 2H-TaSe$_2$ diffraction patterns, with additional super-lattice reflexes, emerging at about 120\:K and changing at about 90\:K.}
  \label{fig:Temperature-series-of}
\end{figure}

As a first working test sample of the cryostat plug-in, we used 2H-TaSe$_2$ possessing a well-documented charge density wave phase transition. In the charge density wave state, an either commensurate or incommensurate super-structure emerges below about 90\:K or 122\:K, respectively \cite{Koenig2013}. The wave vector of the commensurate superstructure is given by $q=\frac{1}{3}q_{0}$, with $q_{0}$ being the absolute value of the wave vector of the basic hexagonal lattice of TaSe$_2$. The wave vector of the incommensurate super-structure is given by $q=\frac{1}{3}\left(1-\delta\right)q_{0}$, with $\delta$ switching to 0.025, at 90\:K and 122\:K, respectively \cite{Moncton1975}. Figure\:\ref{fig:Temperature-series-of} shows a series of diffraction patterns taken at different temperatures. At 130\:K, we only observe the hexagonal diffraction pattern typical for the projection along the $c$-axis of TaSe$_2$. The additional reflexes with the same modulus of the $q$ vector are stemming from differently oriented grains of the crystal due to a slight drift of the specimen during the acquisition of the temperature series. The emerging charge density wave in 2H-TaSe$_2$ suggests that the temperature calibration of the system is working properly, with an accuracy of at least $\pm 2$\:K, even if a temperature gradient between the thermometer and the specimen is taken into account. Because the temperature setting time depends on the absolute temperature, on the heat conductivity and heat capacity of the used TEM grid and of the sample as well as its size and its position on the grid, we waited several minutes before recording diffraction patterns after changing the temperature. Additionally, there is a possible temperature gradient due to electron beam heating, which is largely dependent on the local sample structure and illumination conditions as well as the absolute temperature.

\section{Summary}

We set up a (scanning) transmission electron microscope with a large experimental space accessible from different directions through several large standard ports with an inner diameter of 63\:mm, called \microscope . The resolving power of the microscope is about 3\:nm, where the limiting factor is an insufficient shielding from magnetic stray fields. The optical information limit of this special setup without external disturbances is estimated to 1.3\:nm.

A first working experimental plug-in is a continuous-flow lHe cryostat permitting precisely controllable sample temperatures in a range between 6.5\:K and 400\:K held for several days. Additionally, the plug-in offers four cooled electrical terminals and space for further additions like, \emph{e.\,g.}, mobile probers. The cryostat plug-in is mechanical drift free and vibration free but induces a further reduction of the system's resolving power to 5\:nm most likely due to channeling of magnetic stray fields.

\begin{acknowledgments}
We are greatly indebted to Hannes Lichte for his constant scientific and financial support of this project, without which it would not have been successful. His contribution particularly includes invaluable advice on the optical and vacuum setup, the stray field issues and, most notably, a never-failing inspiration in walking off the beaten tracks. 
This project has received funding from the European Research Council (ERC) under the European Unions Horizon 2020 research and innovation programme (grant agreement No 715620). 
\end{acknowledgments}

\end{document}

%% file: figures/cetcor/cetcor.tex
\definecolor{cb4b6b9}{RGB}{180,182,185}
\definecolor{c211e1f}{RGB}{33,30,31}
\definecolor{c77797b}{RGB}{119,121,123}
\def\y1{0.5}
\begin{tikzpicture}[y=0.80pt, x=0.80pt, yscale=-\y1, xscale=0.8, inner sep=0pt, outer sep=0pt]
	\begin{scope}[cm={{2.062,0.0,0.0,-2.062,(-129.99548,1482.0161)}}]
	
\path[fill=cb4b6b9,nonzero rule,pattern=north west lines,pattern color=lightgray] (280.9290,626.1980) .. controls					
	(280.9290,618.7370) and (252.3650,612.6880) .. (229.4670,612.6880) .. controls
	(206.5680,612.6880) and (178.0060,618.7370) .. (178.0060,626.1980) .. controls
	(178.0060,633.6590) and (206.5680,639.7080) .. (229.4670,639.7080) .. controls
	(252.3650,639.7080) and (280.9290,633.6590) .. (280.9290,626.1980);
\path[draw=c211e1f,line join=miter,line cap=butt,miter limit=4.00,line width=0.400pt] 
(280.9290,626.1980) .. controls (280.9290,618.7370) and
(252.3650,612.6880) .. (229.4670,612.6880) .. controls (206.5680,612.6880) and
(178.0060,618.7370) .. (178.0060,626.1980) .. controls (178.0060,633.6590) and
(206.5680,639.7080) .. (229.4670,639.7080) .. controls (252.3650,639.7080) and
(280.9290,633.6590) .. (280.9290,626.1980) -- cycle;

\path[fill=cb4b6b9,nonzero rule] (283.7750,586.0280) .. controls
(283.7750,582.2590) and (248.4730,579.2040) .. (229.5980,579.2040) .. controls
(210.7230,579.2040) and (175.4200,582.2590) .. (175.4200,586.0280) .. controls
(175.4200,589.7960) and (210.7230,592.8500) .. (229.5980,592.8500) .. controls
(248.4730,592.8500) and (283.7750,589.7960) .. (283.7750,586.0280);
\path[draw=c211e1f,line join=miter,line cap=butt,miter limit=4.00,line
width=0.400pt] (283.7750,586.0280) .. controls (283.7750,582.2590) and
(248.4730,579.2040) .. (229.5980,579.2040) .. controls (210.7230,579.2040) and
(175.4200,582.2590) .. (175.4200,586.0280) .. controls (175.4200,589.7960) and
(210.7230,592.8500) .. (229.5980,592.8500) .. controls (248.4730,592.8500) and
(283.7750,589.7960) .. (283.7750,586.0280) -- cycle;
\path[fill=cb4b6b9,nonzero rule] (283.0290,532.5260) .. controls
(283.0290,528.8980) and (248.7270,525.9580) .. (229.8510,525.9580) .. controls
(210.9750,525.9580) and (175.6740,528.8980) .. (175.6740,532.5260) .. controls
(175.6740,536.1530) and (210.9750,539.0930) .. (229.8510,539.0930) .. controls
(248.7270,539.0930) and (283.0290,536.1530) .. (283.0290,532.5260);
\path[draw=c211e1f,line join=miter,line cap=butt,miter limit=4.00,line
width=0.400pt] (283.0290,532.5260) .. controls (283.0290,528.8980) and
(248.7270,525.9580) .. (229.8510,525.9580) .. controls (210.9750,525.9580) and
(175.6740,528.8980) .. (175.6740,532.5260) .. controls (175.6740,536.1530) and
(210.9750,539.0930) .. (229.8510,539.0930) .. controls (248.7270,539.0930) and
(283.0290,536.1530) .. (283.0290,532.5260) -- cycle;
\path[fill=cb4b6b9,nonzero rule] (283.9060,472.3160) .. controls
(283.9060,468.6220) and (248.6320,465.6260) .. (229.7890,465.6260) .. controls
(210.9470,465.6260) and (175.6720,468.6220) .. (175.6720,472.3160) .. controls
(175.6720,476.0110) and (210.9470,479.0060) .. (229.7890,479.0060) .. controls
(248.6320,479.0060) and (283.9060,476.0110) .. (283.9060,472.3160);
\path[draw=c211e1f,line join=miter,line cap=butt,miter limit=4.00,line
width=0.400pt] (283.9060,472.3160) .. controls (283.9060,468.6220) and
(248.6320,465.6260) .. (229.7890,465.6260) .. controls (210.9470,465.6260) and
(175.6720,468.6220) .. (175.6720,472.3160) .. controls (175.6720,476.0110) and
(210.9470,479.0060) .. (229.7890,479.0060) .. controls (248.6320,479.0060) and
(283.9060,476.0110) .. (283.9060,472.3160) -- cycle;
\path[fill=cb4b6b9,nonzero rule] (283.9060,432.8440) .. controls
(283.9060,429.1500) and (248.6320,426.1550) .. (229.7890,426.1550) .. controls
(210.9470,426.1550) and (175.6720,429.1500) .. (175.6720,432.8440) .. controls
(175.6720,436.5390) and (210.9470,439.5350) .. (229.7890,439.5350) .. controls
(248.6320,439.5350) and (283.9060,436.5390) .. (283.9060,432.8440);
\path[draw=c211e1f,line join=miter,line cap=butt,miter limit=4.00,line
width=0.400pt] (283.9060,432.8440) .. controls (283.9060,429.1500) and
(248.6320,426.1550) .. (229.7890,426.1550) .. controls (210.9470,426.1550) and
(175.6720,429.1500) .. (175.6720,432.8440) .. controls (175.6720,436.5390) and
(210.9470,439.5350) .. (229.7890,439.5350) .. controls (248.6320,439.5350) and
(283.9060,436.5390) .. (283.9060,432.8440) -- cycle;
\path[fill=cb4b6b9,nonzero rule,pattern=north west lines,pattern color=lightgray] (283.9060,391.8650) .. controls
(283.9060,388.1700) and (248.6320,385.1750) .. (229.7890,385.1750) .. controls
(210.9470,385.1750) and (175.6720,388.1700) .. (175.6720,391.8650) .. controls
(175.6720,395.5590) and (210.9470,398.5550) .. (229.7890,398.5550) .. controls
(248.6320,398.5550) and (283.9060,395.5590) .. (283.9060,391.8650);

\node[color=red!80!black, anchor=base west, yshift=-8*\y1, xshift=5] at (286.9230,391.8650) {ADL};
\node[color=red!80!black, anchor=base west, yshift=6*\y1, xshift=5] at (286.9230,406.0820) {HP2};
\node[color=red!80!black, anchor=base west, yshift=8*\y1, xshift=5] at (286.9230,486.2910) {HP1};
\node[color=red!80!black, anchor=base west, yshift=-8*\y1, xshift=5] at (286.9230,432.8440) {TL22};
\node[color=red!80!black, anchor=base west, yshift=-8*\y1, xshift=5] at (286.9230,472.3160) {Tl21};
\node[color=red!80!black, anchor=base west, yshift=-8*\y1, xshift=5] at (286.9230,532.5260) {TL12};
\node[color=red!80!black, anchor=base west, yshift=-8*\y1, xshift=5] at (286.9230,586.0280) {TL11};

\node[color=red!80!black, anchor=base west, yshift=0, xshift=5] at (286.9230,322.0650) {SA plane};
\node[color=red!80!black, anchor=base west, yshift=0, xshift=5] at (286.9230,626.0810) {specimen};


\path[draw=c211e1f,line join=miter,line cap=butt,miter limit=4.00,line
width=0.400pt] (283.9060,391.8650) .. controls (283.9060,388.1700) and
(248.6320,385.1750) .. (229.7890,385.1750) .. controls (210.9470,385.1750) and
(175.6720,388.1700) .. (175.6720,391.8650) .. controls (175.6720,395.5590) and
(210.9470,398.5550) .. (229.7890,398.5550) .. controls (248.6320,398.5550) and
(283.9060,395.5590) .. (283.9060,391.8650) -- cycle;
\path[fill=c77797b,nonzero rule] (175.7070,486.2910) -- (283.8730,486.2910) --
(283.8730,498.3750) -- (175.7070,498.3750) -- (175.7070,486.2910) -- cycle;
\path[draw=c211e1f,line join=miter,line cap=butt,miter limit=4.00,line
width=0.400pt] (283.8730,486.2910) -- (175.7070,486.2910) --
(175.7070,498.3750) -- (283.8730,498.3750) -- (283.8730,486.2910) -- cycle;
\path[fill=c77797b,nonzero rule] (175.7070,406.0820) -- (283.8730,406.0820) --
(283.8730,418.1660) -- (175.7070,418.1660) -- (175.7070,406.0820) -- cycle;
\path[draw=c211e1f,line join=miter,line cap=butt,miter limit=4.00,line
width=0.400pt] (283.8730,406.0820) -- (175.7070,406.0820) --
(175.7070,418.1660) -- (283.8730,418.1660) -- (283.8730,406.0820) --
cycle(220.2560,587.3940) -- (215.4649,587.3940) -- (215.4649,602.3210) --
(220.2560,602.3210) -- (220.2560,587.3940) -- cycle(215.4650,602.3210) --
(220.2560,587.3940)(215.5760,587.3940) --
(220.2560,602.3210)(243.8210,587.3940) -- (239.0309,587.3940) --
(239.0309,602.3210) -- (243.8210,602.3210) -- (243.8210,587.3940) --
cycle(239.0310,602.3210) -- (243.8210,587.3940)(239.1420,587.3940) --
(243.8210,602.3210)(217.5820,546.8440) -- (212.7910,546.8440) --
(212.7910,555.5315) -- (217.5820,555.5315) -- (217.5820,546.8440) --
cycle(212.7910,555.5310) -- (217.5820,546.8440)(212.9020,546.8440) --
(217.5820,555.5310)(246.3240,546.8440) -- (241.5340,546.8440) --
(241.5340,555.5315) -- (246.3240,555.5315) -- (246.3240,546.8440) --
cycle(241.5340,555.5310) -- (246.3240,546.8440)(241.6450,546.8440) --
(246.3240,555.5310)(217.5820,488.0800) -- (212.7910,488.0800) --
(212.7910,496.7670) -- (217.5820,496.7670) -- (217.5820,488.0800) --
cycle(212.7910,496.7670) -- (217.5820,488.0800)(212.9020,488.0800) --
(217.5820,496.7670)(246.3240,488.0800) -- (241.5340,488.0800) --
(241.5340,496.7670) -- (246.3240,496.7670) -- (246.3240,488.0800) --
cycle(241.5340,496.7670) -- (246.3240,488.0800)(241.6450,488.0800) --
(246.3240,496.7670)(217.5820,461.2320) -- (212.7910,461.2320) --
(212.7910,469.9190) -- (217.5820,469.9190) -- (217.5820,461.2320) --
cycle(212.7910,469.9190) -- (217.5820,461.2320)(212.9020,461.2320) --
(217.5820,469.9190)(246.3240,461.2320) -- (241.5340,461.2320) --
(241.5340,469.9190) -- (246.3240,469.9190) -- (246.3240,461.2320) --
cycle(241.5340,469.9190) -- (246.3240,461.2320)(241.6450,461.2320) --
(246.3240,469.9190)(217.5820,435.0540) -- (212.7910,435.0540) --
(212.7910,443.7420) -- (217.5820,443.7420) -- (217.5820,435.0540) --
cycle(212.7910,443.7420) -- (217.5820,435.0540)(212.9020,435.0540) --
(217.5820,443.7420)(246.3240,435.0540) -- (241.5340,435.0540) --
(241.5340,443.7420) -- (246.3240,443.7420) -- (246.3240,435.0540) --
cycle(241.5340,443.7420) -- (246.3240,435.0540)(241.6450,435.0540) --
(246.3240,443.7420)(217.5820,407.7510) -- (212.7910,407.7510) --
(212.7910,416.4390) -- (217.5820,416.4390) -- (217.5820,407.7510) --
cycle(212.7910,416.4390) -- (217.5820,407.7510)(212.9020,407.7510) --
(217.5820,416.4390)(246.3240,407.7510) -- (241.5340,407.7510) --
(241.5340,416.4390) -- (246.3240,416.4390) -- (246.3240,407.7510) --
cycle(241.5340,416.4390) -- (246.3240,407.7510)(241.6450,407.7510) --
(246.3240,416.4390)(217.5820,370.6620) -- (212.7910,370.6620) --
(212.7910,379.3490) -- (217.5820,379.3490) -- (217.5820,370.6620) --
cycle(212.7910,379.3490) -- (217.5820,370.6620)(212.9020,370.6620) --
(217.5820,379.3490)(246.3240,370.6620) -- (241.5340,370.6620) --
(241.5340,379.3490) -- (246.3240,379.3490) -- (246.3240,370.6620) --
cycle(241.5340,379.3490) -- (246.3240,370.6620)(241.6450,370.6620) --
(246.3240,379.3490)(217.5820,342.5840) -- (212.7910,342.5840) --
(212.7910,351.2720) -- (217.5820,351.2720) -- (217.5820,342.5840) --
cycle(212.7910,351.2720) -- (217.5820,342.5840)(212.9020,342.5840) --
(217.5820,351.2720)(246.3240,342.5840) -- (241.5340,342.5840) --
(241.5340,351.2720) -- (246.3240,351.2720) -- (246.3240,342.5840) --
cycle(241.5340,351.2720) -- (246.3240,342.5840)(241.6450,342.5840) --
(246.3240,351.2720)(229.4670,639.7080) -- (229.4670,306.5820);

\node[color=red!80!black, anchor=base west, yshift=-12*\y1, xshift=-30] at (166.0970,379.3490) {ISh};
\node[color=red!80!black, anchor=base west, yshift=-12*\y1, xshift=-30] at (166.0970,416.4390) {DPH2};
\node[color=red!80!black, anchor=base west, yshift=-12*\y1, xshift=-30] at (166.0970,443.7420) {DP22};
\node[color=red!80!black, anchor=base west, yshift=-12*\y1, xshift=-30] at (166.0970,469.9190) {DP21};
\node[color=red!80!black, anchor=base west, yshift=-12*\y1, xshift=-30] at (166.0970,496.7670) {DPH1};
\node[color=red!80!black, anchor=base west, yshift=-12*\y1, xshift=-30] at (166.0970,555.5310) {DP12};

\node[color=red!80!black, anchor=base west, yshift=0, xshift=-31.5] at (166.8500,626.0810) {OL};
\node[color=red!80!black, anchor=base west, yshift=-3, xshift=-30] at (166.0970,338.8890) {DStig};
\node[color=red!80!black, anchor=base west, yshift=-3, xshift=-30] at (166.0970,355.7310) {DSh};
\node[color=red!80!black, anchor=base west, yshift=-3, xshift=-30] at (166.0970,582.0560) {DP11};
\node[color=red!80!black, anchor=base west, yshift=-3, xshift=-30] at (166.0970,594.7750) {QPol};
\node[color=red!80!black, anchor=base west, yshift=-3, xshift=-30] at (166.0970,609.0010) {HPol};

\path[fill=c211e1f,nonzero rule] (229.4670,303.5640) .. controls
(228.9400,304.9840) and (228.0410,306.7460) .. (227.0900,307.8370) --
(229.4670,306.9770) -- (231.8440,307.8370) .. controls (230.8930,306.7460) and
(229.9930,304.9840) .. (229.4670,303.5640);
\path[draw=c211e1f,line join=miter,line cap=butt,miter limit=4.00,line
width=0.400pt] 
(173.0060,322.0650)--(285.9290,322.0650)    
(173.0060,626.0810) -- (285.9290,626.0810)  
(227.7460,632.2420) -- (231.4110,632.2420)
(227.7460,620.1720) -- (231.4110,620.1720)
(165.9560,355.7310) -- (168.9040,355.7310) -- (168.9040,338.8890) -- (166.0970,338.8890)
(168.9040,347.1690) -- (171.9910,347.1690)
(165.9560,609.0010) -- (168.9040,609.0010) -- (168.9040,582.0560) -- (166.0970,582.0560)
(168.9040,594.7750) -- (171.9910,594.7750);

\path[draw=c211e1f,line join=miter,line cap=butt,miter limit=4.00,line
width=0.400pt,rounded corners=5pt, dashed, color=black]
(229.4670,626.0810) -- (275,620.0810) -- (275,585.9140) -- (202.8500,532.4150)--
(202.8500,472.0820) -- (256.1830,432.7480) -- (256.5170,391.9150) -- (229.4670,322.0650);

\path[draw=c211e1f,line join=miter,line cap=butt,miter limit=4.00,line
width=0.400pt,rounded corners=5pt, dashed, color=red!80!black]
(229.4670,620.1720) -- (265,586.0810) -- (265,532.0820) -- (210.6840,471.9150) --
(210.6840,432.5820) -- (249.1840,391.5820) -- (270.8500,322.0650);

\path[draw=c211e1f,line join=miter,line cap=butt,miter limit=4.00,line
width=0.400pt, color=black,rounded corners=5pt]
(250,626.0810) -- (250,585.9140) -- (219,532.4150)--
(180,472.0820) -- (260,432.7480) -- (275,322.0650);

\path[draw=c211e1f,line join=miter,line cap=butt,miter limit=4.00,line
width=0.400pt, color=red!80!black,rounded corners=5pt]
(229.4670,626.0810)-- (275,586.0810) -- (265,532.0820) --  (245,471.9150) -- 
(180,432.5820) -- (229.4670,322.0650);


\node[color=red!80!black, anchor=base west, yshift=-5*\y1, xshift=-23] at (231.4110,632.2420) {ffp};
\node[color=red!80!black, anchor=base west, yshift=-5*\y1, xshift=-23] at (231.4110,620.1720) {bfp};

\end{scope}
\end{tikzpicture}

%% file: figures/chamber/chamber.tex
\pgfmathsetlength{\imagewidth}{\columnwidth}
\pgfmathsetlength{\imagescale}{\imagewidth/1024}

\begin{tikzpicture}[x=\imagescale,y=-\imagescale]
      \sffamily
  \node[anchor=north west,inner sep=0pt,outer sep=0pt] at (0,0) {\includegraphics[width=\imagewidth]{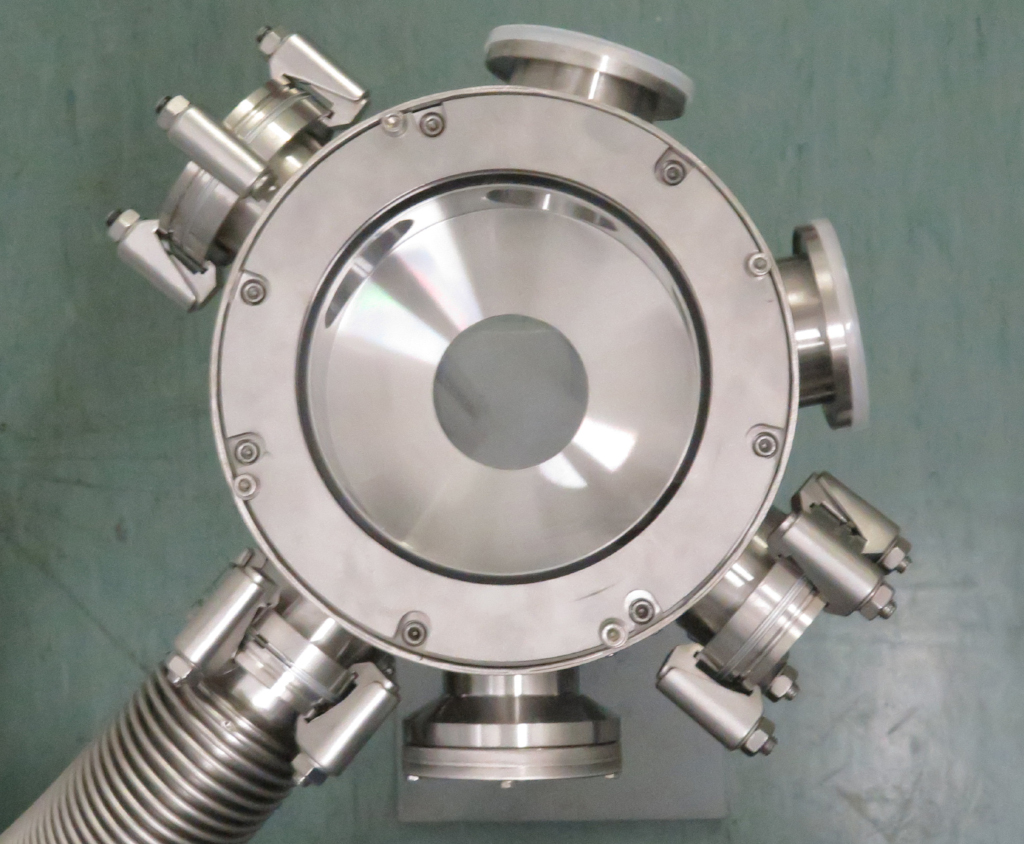}};
  \draw[ultra thick,white] (772,800) node [anchor=south west] {100\:mm} -- (1000,800);
  \draw[thick,color=orange] (505,380) circle (2.49cm); 
  \node[orange] at (30,370) [anchor=west] {permalloy} ;
  \node[orange] at (30,410) [anchor=west] {shielding} ;
  \node[yellow] at (850,100) [anchor=north] {ISO-K\:63 ports} ;
    \node[yellow] at (512,790) [anchor=north] {IGP port} ;
\end{tikzpicture}%

%% file: figures/cryo_layout/cryo_layout.tex
    \pgfmathsetlength{\imagewidth}{\columnwidth}
    \pgfmathsetlength{\imagescale}{\imagewidth/1024}

    \begin{tikzpicture}[x=\imagescale,y=-\imagescale]
      \sffamily
        \node[anchor=north west,inner sep=0pt,outer sep=0pt] at (0,0)
        {\includegraphics[width=\columnwidth]{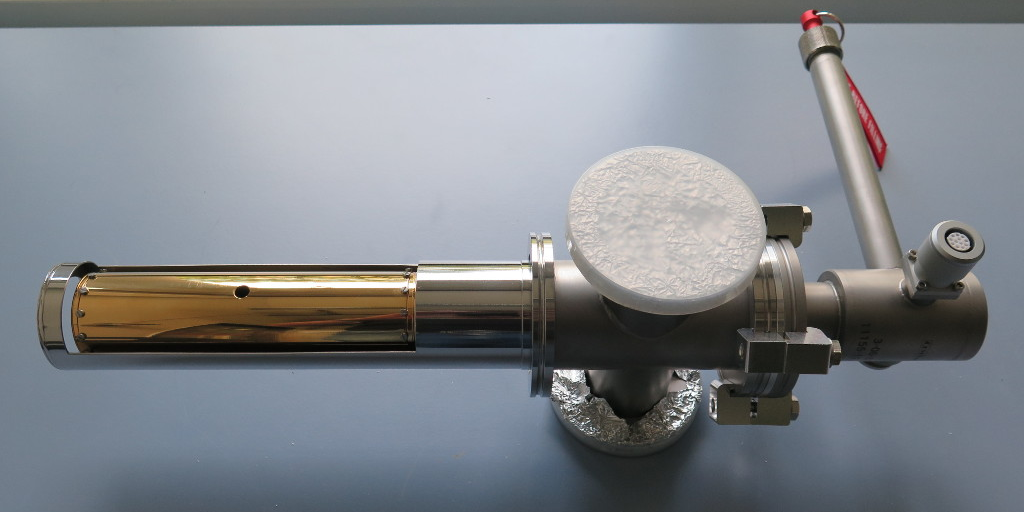}};
        \node[anchor=north west] at (0,0) {(a)};
        \draw [->, thick,red] (200,100) node[anchor=south]{electron beam hole} -- (240,280) ;
        \draw [->, thick] (450,150) node[anchor=south]{cooled radiation shield} -- (350,300) ;
        \draw [->, thick] (700,50) node[anchor=east]{liquid helium} -- (780,40) ;
        \draw [->, thick,yellow] (880,450) node[anchor=north]{helium exhaust} -- (880,380) ;
        \draw [->, thick,yellow] (450,450) node[anchor=north east]{auxiliary feedthrough ports} -- (660,240) ;
        \draw [->, thick,yellow] (460,460) -- (600,390) ;
      \end{tikzpicture}%
\vskip1ex
      \begin{tikzpicture}[x=\imagescale,y=-\imagescale]
      \sffamily
        \node[anchor=north west,inner sep=0pt,outer sep=0pt] at (0,0)
        {\includegraphics[width=\columnwidth]{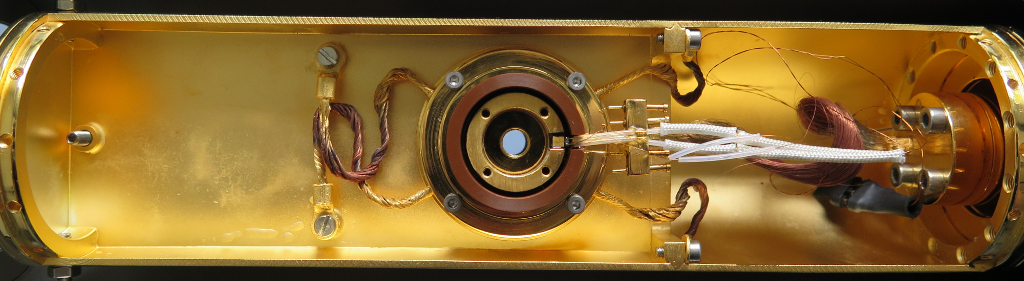}};
        \node[anchor=north west] at (0,-200) {(b)};
        \draw [->, thick] (350,-100) node[anchor=south]{ring-shaped sample mount on stage} -- (500,110) ;
        \draw [->, thick] (650,-50) node[anchor=south]{4 cooled electrical terminals} -- (630,110) ;
        \draw [->, thick] (900,-100) node[anchor=south]{cold finger} -- (970,110) ;
        \draw [->, thick] (500,350) node[anchor=north]{space for additions} -- (200,200) ;
        \draw [->, thick] (524,350)  -- (800,200) ;
        \draw [ultra thick] (750,350) node[anchor=south west]{50 mm} -- (1000,350) ;
      \end{tikzpicture}%

%% file: figures/cryo_resolution/cryo_resolution.tex
\pgfmathsetlength{\imagewidth}{.49\columnwidth}
\pgfmathsetlength{\imagescale}{\imagewidth/1024}
      \begin{tikzpicture}[x=\imagescale,y=-\imagescale]
      \sffamily
        \node[anchor=north west,inner sep=0pt,outer sep=0pt] at (0,0)
        {\includegraphics[width=\imagewidth]{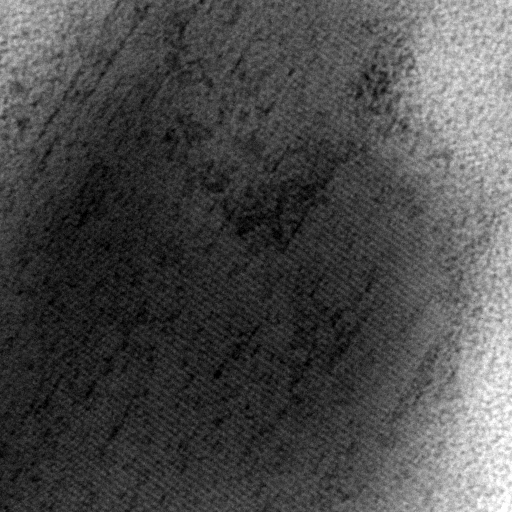}};
        \draw[ultra thick,white] (24,1000) node [anchor=south west] {100\:nm} -- (320,1000);
        \node at (0,0) [anchor=north west,white] {(a)} ;

      \end{tikzpicture}%
      \hfill
      \begin{tikzpicture}[x=\imagescale,y=-\imagescale]
      \sffamily
        \node[anchor=north west,inner sep=0pt,outer sep=0pt] at (0,0)
        {\includegraphics[width=\imagewidth]{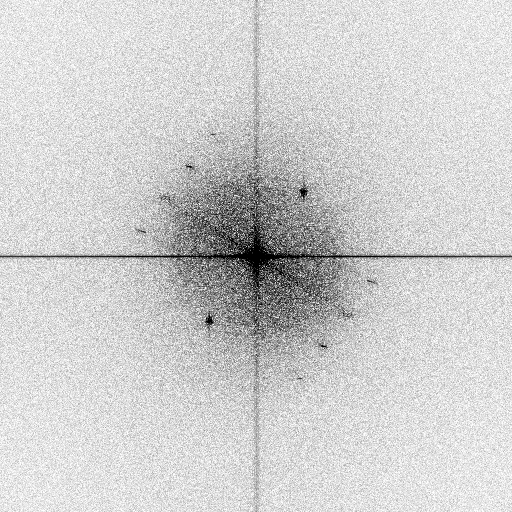}};
        \draw[thick,dashed] (512,512) circle (275);
        \node at (512,237) [anchor=south west] {5\:nm} ;
        \draw[ultra thick] (24,1000) node [anchor=south west] {0.2\:nm$^{\text{--1}}$} -- (300,1000);
        \node at (0,0) [anchor=north west] {(b)} ;
      \end{tikzpicture}%

%% file: figures/imagespread/imagespread.tex
 \begin{tikzpicture}[tight background,rotate=-90]

    \coordinate (O) at (0,0) ;
    
    \draw[ultra thick] (-2,-0.5) -- (-2,0.5) ;
    \draw[red!80!black,ultra thick] (4,-1) -- (4,1) ;
    \draw[ultra thick] (4,-2) -- (4,0) ;
    \draw[red!20!blue, ultra thick] (-2,0) -- (-2,-1); 
	\node[below, text width=30,red!20!blue] at (-2.4,-1.7) {``shifted''};
	\node[below, text width=30,red!20!blue] at (-2.1,-1.55) {object};
    \node[above] at (-1.92,0.5) {object};
    \node[above] at (4.2,1.5) {image};
    \node[above,opacity=0.4] at (-1.045,1.35) {B};
    \node[above,opacity=0.4] at (-0.0,1) {$\bigotimes$};
    \node[above,opacity=0.4] at (-0.5,1) {$\bigotimes$};
    \node[above,opacity=0.4] at (-1,1) {$\bigotimes$};
    \node[above,opacity=0.4] at (-1.5,1) {$\bigotimes$};

    \draw[dash pattern=on5pt off3pt] (-2,0) -- (4,0) ;
    \draw[<->] (0,-2) -- (0,2) ;
    \node[right] at (-0.12,2) {TL11};

    \draw[red!80!black] (-2,0.5) -- (0,0.5);
    \draw[red!80!black] (-2,0.5) -- (0,0.75);
    \draw[red!80!black] (-2,-0.5) -- (0,-0.5);
    \draw[red!80!black] (-2,-0.5) -- (0,-0.25);
    
    \draw[thick] (-2,0.5) parabola (0,1);
    \draw[thick] plot[smooth,domain=-2:0] (\x, {0.75+0.25/2*\x+0.125*(\x+2)*(\x+2)});
    \draw[thick] (-2,-0.5) parabola (0,0);
    \draw[thick] plot[smooth,domain=-2:0] (\x, {-0.25+0.25/2*\x+0.125*(\x+2)*(\x+2)});
    
    \draw[red!20!blue] (-2,0) -- (0,1);
    \draw[red!20!blue] (-2,0) -- (0,1.25);
    \draw[red!20!blue] (-2,-1) -- (0,0.25);
    \draw[red!20!blue] (-2,-1) -- (0,0);
    
    \draw[thick] (0,1) -- (4,0);
    \draw[thick] (0,1.25) -- (4,0);
    \draw[thick] (0,0.25) -- (4,-2);
    \draw[thick] (0,0) -- (4,-2);
    
    \draw[red!80!black] (0,0.5) -- (4,-1);
    \draw[red!80!black] (0,0.75) -- (4,-1);
    \draw[red!80!black] (0,-0.5) -- (4,1);
    \draw[red!80!black] (0,-0.25) -- (4,1);


\end{tikzpicture}

%% file: figures/ctf/fs_dampening.tex
%
%
%
\begin{tikzpicture}[tight background]

  \pgfplotsset{set layers}
  \begin{axis}[
      axis y line*=right,
      axis x line*=top,
      xlabel={distance / nm},
      xmin=0, xmax=1.6,
      ymin=-0.3, ymax=1.2,
      ytick=\empty,
      xtick={0.2,0.3333,0.5,0.6667,1},
      xticklabels={5,3,2,1.5,1},
      width=\columnwidth,
      height=.5\columnwidth
    ]
  \end{axis}
\begin{axis}[
      axis y line*=left,
      axis x line*=bottom,
xlabel={spatial frequency / nm$^{-1}$},
ylabel={contrast transfer},
      xtick={0,0.2,0.4,0.6,0.8,1.0,1.2,1.4},
      ytick={0,1},
xmin=0, xmax=1.6,
ymin=-0.3, ymax=1.2,
axis on top,
width=\columnwidth,
height=.5\columnwidth
]
  \node[anchor=south west,blue] at (0.3,0.135) {\footnotesize image spread};
  \node[anchor=south west,green!50.0!black] at (0.77,0.135) {\footnotesize temporal coherence};
  \node[anchor=north west,green!50.0!black] at (0.9,1) {\footnotesize spatial coherence};
  
\node[anchor=west] at (0.1,0.01) {noise};
\addplot [fill=black, draw=none, fill opacity=0.3]
table {%
0 0.135
1.6 0.135
}\closedcycle;
\addplot [fill=black, draw=none, fill opacity=0.3]
table {%
0 -0.135
1.6 -0.135
}\closedcycle;
\addplot [dashed,black]
table {%
0 0
1.6 0
};
\addplot [blue]
table {%
0 1
0.00804020100502513 0.999052256310181
0.0160804020100503 0.996214181679639
0.0241206030150754 0.991501200882024
0.0321608040201005 0.984938873906805
0.0402010050251256 0.976562675788678
0.0482412060301508 0.966417691902004
0.0562814070351759 0.9545582321948
0.064321608040201 0.941047368746599
0.0723618090452261 0.925956401883939
0.0804020100502513 0.909364260864841
0.0884422110552764 0.891356845839078
0.0964824120603015 0.872026318395324
0.104522613065327 0.851470348511687
0.112562814070352 0.829791326126697
0.120603015075377 0.807095545838973
0.128643216080402 0.783492373422832
0.136683417085427 0.759093402912822
0.144723618090452 0.734011612963319
0.152763819095477 0.708360531032083
0.160804020100503 0.682253413673071
0.168844221105528 0.655802450859226
0.176884422110553 0.629118001797349
0.184924623115578 0.602307869152735
0.192964824120603 0.575476617980315
0.201005025125628 0.548724944972058
0.209045226130653 0.52214910288835
0.217085427135678 0.495840384255888
0.225125628140704 0.4698846675984
0.233165829145729 0.444362028631614
0.241206030150754 0.419346418012732
0.249246231155779 0.39490540639936
0.257286432160804 0.371099996755224
0.265326633165829 0.347984503051147
0.273366834170854 0.325606493760196
0.281407035175879 0.304006797845055
0.289447236180905 0.283219570292097
0.29748743718593 0.263272413667496
0.305527638190955 0.244186551662244
0.31356783919598 0.225977050159709
0.321608040201005 0.208653081004916
0.32964824120603 0.192218223380903
0.337688442211055 0.176670797505134
0.34572864321608 0.162004225247175
0.353768844221106 0.148207412235833
0.361809045226131 0.135265146066554
0.369849246231156 0.12315850533377
0.377889447236181 0.111865274392945
0.385929648241206 0.101360358997218
0.393969849246231 0.091616198246921
0.402010050251256 0.0826031686296782
0.410050251256281 0.074289976306405
0.418090452261307 0.0666440342064213
0.426130653266332 0.059631820924986
0.434170854271357 0.0532192188608284
0.442211055276382 0.0473718294819033
0.450251256281407 0.0420552640571561
0.458291457286432 0.0372354086335283
0.466331658291457 0.0328786624643132
0.474371859296482 0.028952149501487
0.482412060301508 0.0254239029456915
0.490452261306533 0.02226302319885
0.498492462311558 0.0194398098824457
0.506532663316583 0.0169258688666284
0.514572864321608 0.0146941954997019
0.522613065326633 0.0127192354331537
0.530653266331658 0.0109769246039751
0.538693467336683 0.0094447100640502
0.546733668341709 0.00810155343702872
0.554773869346734 0.00692791883805681
0.562814070351759 0.00590574711330384
0.570854271356784 0.00501841824706155
0.578894472361809 0.00425070374735008
0.586934673366834 0.00358871075972804
0.594974874371859 0.00301981957682351
0.603015075376884 0.00253261611151731
0.61105527638191 0.00211682078825236
0.619095477386935 0.00176321518307301
0.62713567839196 0.00146356761203811
0.635175879396985 0.00121055873272062
0.64321608040201 0.000997708087481536
0.651256281407035 0.000819302382678508
0.65929648241206 0.000670326167217855
0.667336683417086 0.00054639544883408
0.675376884422111 0.000443694668793328
0.683417085427136 0.000358917346641246
0.691457286432161 0.00028921060709605
0.699497487437186 0.000232123711856262
0.707537688442211 0.0001855606402902
0.715577889447236 0.00014773669477718
0.723618090452261 0.000117139048720101
0.731658291457287 9.2491107586027e-05
0.739698492462312 7.27205152277799e-05
0.747738693467337 5.69306085346072e-05
0.755778894472362 4.4375102392239e-05
0.763819095477387 3.44357731726805e-05
0.771859296482412 2.66029016533238e-05
0.779899497487437 2.04582345026397e-05
0.787939698492462 1.56602263960396e-05
0.795979899497487 1.19313316015819e-05
0.804020100502513 9.04712370910867e-06
0.812060301507538 6.82703433585895e-06
0.820100502512563 5.12651546295638e-06
0.828140703517588 3.83044495126019e-06
0.836180904522613 2.84761024004382e-06
0.844221105527638 2.1061208135228e-06
0.852261306532663 1.54961536962216e-06
0.860301507537689 1.13414445550717e-06
0.868341708542714 8.25623424739832e-07
0.876381909547739 5.97763761145482e-07
0.884422110552764 4.30402997586941e-07
0.892462311557789 3.08164572961641e-07
0.900502512562814 2.19388995809299e-07
0.908542713567839 1.55286627673149e-07
0.916582914572864 1.09270298639045e-07
0.92462311557789 7.64328751190599e-08
0.932663316582915 5.31408832964146e-08
0.94070351758794 3.67204267837551e-08
0.948743718592965 2.52160047154958e-08
0.95678391959799 1.72065187707425e-08
0.964824120603015 1.1665835240243e-08
0.97286432160804 7.85781856703476e-09
0.980904522613065 5.25784831711103e-09
0.988944723618091 3.49453895874159e-09
0.996984924623116 2.30676144448675e-09
1.00502512562814 1.51217108801903e-09
1.01306532663317 9.84324740726274e-10
1.02110552763819 6.36162630248581e-10
1.02914572864322 4.0817135243183e-10
1.03718592964824 2.59963932108523e-10
1.04522613065327 1.6433525330072e-10
1.05326633165829 1.0309688570694e-10
1.06130653266332 6.41810567616242e-11
1.06934673366834 3.96427034103135e-11
1.07738693467337 2.42919591210452e-11
1.08542713567839 1.47656207140926e-11
1.09346733668342 8.90178249489755e-12
1.10150753768844 5.32210897251341e-12
1.10954773869347 3.15512564423027e-12
1.11758793969849 1.85446841883249e-12
1.12562814070352 1.0805255909477e-12
1.13366834170854 6.24030201010982e-13
1.14170854271357 3.57167843038519e-13
1.14974874371859 2.0257025751294e-13
1.15778894472362 1.13829577401866e-13
1.16582914572864 6.33649841910459e-14
1.17386934673367 3.49378410603103e-14
1.18190954773869 1.90779713011311e-14
1.18994974874372 1.03155804223115e-14
1.19798994974874 5.52224336861665e-15
1.20603015075377 2.9263881377706e-15
1.21407035175879 1.53488558310675e-15
1.22211055276382 7.96672847577679e-16
1.23015075376884 4.0914310276737e-16
1.23819095477387 2.07870077225629e-16
1.24623115577889 1.04462299967434e-16
1.25427135678392 5.19166194429916e-17
1.26231155778894 2.55128984261963e-17
1.27035175879397 1.23949881778623e-17
1.278391959799 5.95238055595997e-18
1.28643216080402 2.82499590901983e-18
1.29447236180905 1.32480142486474e-18
1.30251256281407 6.13778679982203e-19
1.3105527638191 2.80880778855369e-19
1.31859296482412 1.26940891192275e-19
1.32663316582915 5.66460166801436e-20
1.33467336683417 2.49541877601071e-20
1.3427135678392 1.08502700505589e-20
1.35075376884422 4.65560800404191e-21
1.35879396984925 1.97090455863613e-21
1.36683417085427 8.23041160267922e-22
1.3748743718593 3.389663492523e-22
1.38291457286432 1.37651778502243e-22
1.39095477386935 5.51070608592741e-23
1.39899497487437 2.17441102106062e-23
1.4070351758794 8.45459097375586e-24
1.41507537688442 3.23866871750291e-24
1.42311557788945 1.22199265360661e-24
1.43115577889447 4.5404903903468e-25
1.4391959798995 1.66100805728373e-25
1.44723618090452 5.98104396889222e-26
1.45527638190955 2.11942831459495e-26
1.46331658291457 7.38917376381923e-27
1.4713567839196 2.53399691371392e-27
1.47939698492462 8.545661465529e-28
1.48743718592965 2.83341196902324e-28
1.49547738693467 9.23404348389303e-29
1.5035175879397 2.95722683250298e-29
1.51155778894472 9.30419761876223e-30
1.51959798994975 2.87517612915976e-30
1.52763819095477 8.72427859395679e-31
1.5356783919598 2.59872940234028e-31
1.54371859296482 7.5970429408465e-32
1.55175879396985 2.17903498536272e-32
1.55979899497487 6.13059488535675e-33
1.5678391959799 1.69137959075812e-33
1.57587939698492 4.57466625865831e-34
1.58391959798995 1.21265136445832e-34
1.59195979899497 3.14954338481328e-35
1.6 8.01253083120821e-36
};
\addplot [green!50.0!black, dashed]
table {%
0 1
0.00804020100502513 0.999999982917366
0.0160804020100503 0.999999726677891
0.0241206030150754 0.999998616307593
0.0321608040201005 0.999995626855225
0.0402010050251256 0.999989323410663
0.0482412060301508 0.999977861151233
0.0562814070351759 0.99995898543657
0.064321608040201 0.999930031978473
0.0723618090452261 0.999887927118097
0.0804020100502513 0.999829188248673
0.0884422110552764 0.999749924427789
0.0964824120603015 0.999645837229093
0.104522613065327 0.99951222188901
0.112562814070352 0.999343968809822
0.120603015075377 0.999135565486058
0.128643216080402 0.998881098926683
0.136683417085427 0.99857425865098
0.144723618090452 0.998208340341251
0.152763819095477 0.997776250240486
0.160804020100503 0.997270510387959
0.168844221105528 0.996683264790187
0.176884422110553 0.996006286628824
0.184924623115578 0.995230986610815
0.192964824120603 0.994348422569368
0.201005025125628 0.993349310427013
0.209045226130653 0.992224036634076
0.217085427135678 0.990962672197272
0.225125628140704 0.989554988413646
0.233165829145729 0.987990474424769
0.241206030150754 0.986258356704708
0.249246231155779 0.984347620592882
0.257286432160804 0.982247033979201
0.265326633165829 0.979945173243959
0.273366834170854 0.977430451548477
0.281407035175879 0.974691149564621
0.289447236180905 0.971715448721669
0.29748743718593 0.968491467037732
0.305527638190955 0.965007297589723
0.31356783919598 0.961251049660839
0.321608040201005 0.95721089258739
0.32964824120603 0.952875102307707
0.337688442211055 0.94823211059463
0.34572864321608 0.943270556929683
0.353768844221106 0.937979342951601
0.361809045226131 0.932347689384279
0.369849246231156 0.926365195319574
0.377889447236181 0.920021899698862
0.385929648241206 0.913308344803815
0.393969849246231 0.906215641531842
0.402010050251256 0.898735536195117
0.410050251256281 0.890860478544448
0.418090452261307 0.882583690680635
0.426130653266332 0.87389923647687
0.434170854271357 0.864802091096484
0.442211055276382 0.855288210151403
0.450251256281407 0.845354598008608
0.458291457286432 0.834999374715167
0.466331658291457 0.824221840977699
0.474371859296482 0.813022540600034
0.482412060301508 0.801403319754058
0.490452261306533 0.789367382434005
0.498492462311558 0.776919341424461
0.506532663316583 0.764065264097882
0.514572864321608 0.750812712349269
0.522613065326633 0.737170775974407
0.530653266331658 0.723150098804664
0.538693467336683 0.708762896926278
0.546733668341709 0.694022968336038
0.554773869346734 0.678945693418807
0.562814070351759 0.663548025675908
0.570854271356784 0.647848472187317
0.578894472361809 0.631867063355112
0.586934673366834 0.615625311550735
0.594974874371859 0.599146158374296
0.603015075376884 0.582453910329984
0.61105527638191 0.565574162827275
0.619095477386935 0.548533712532253
0.62713567839196 0.531360458216087
0.635175879396985 0.514083290377454
0.64321608040201 0.496731970050962
0.651256281407035 0.479336997352989
0.65929648241206 0.461929470457861
0.667336683417086 0.444540935839068
0.675376884422111 0.427203230750047
0.683417085427136 0.409948319054609
0.691457286432161 0.392808121646037
0.699497487437186 0.375814342813607
0.707537688442211 0.35899829402341
0.715577889447236 0.342390716674248
0.723618090452261 0.326021605466793
0.731658291457287 0.309920034082697
0.739698492462312 0.294113984907977
0.747738693467337 0.278630184549891
0.755778894472362 0.263493946887175
0.763819095477387 0.248729025358815
0.771859296482412 0.234357476135744
0.779899497487437 0.220399533732802
0.787939698492462 0.206873500505146
0.795979899497487 0.193795651334966
0.804020100502513 0.181180154652094
0.812060301507538 0.169039010747858
0.820100502512563 0.157382008137684
0.828140703517588 0.146216698507541
0.836180904522613 0.135548390545632
0.844221105527638 0.125380162717705
0.852261306532663 0.115712894796091
0.860301507537689 0.106545317703468
0.868341708542714 0.0978740809871089
0.876381909547739 0.089693837002467
0.884422110552764 0.0819973406611183
0.892462311557789 0.0747755633916827
0.900502512562814 0.068017819777508
0.908542713567839 0.0617119051754111
0.916582914572864 0.0558442424888538
0.92462311557789 0.0504000361692603
0.932663316582915 0.0453634314527332
0.94070351758794 0.040717676807471
0.948743718592965 0.0364452875701903
0.95678391959799 0.0325282087875075
0.964824120603015 0.0289479753494058
0.97286432160804 0.0256858676047241
0.980904522613065 0.022723060780432
0.988944723618091 0.0200407666840319
0.996984924623116 0.0176203663479166
1.00502512562814 0.0154435324716061
1.01306532663317 0.0134923407278706
1.02110552763819 0.011749369216957
1.02914572864322 0.010197785574571
1.03718592964824 0.00882142145907678
1.04522613065327 0.00760483435690444
1.05326633165829 0.00653335684807909
1.06130653266332 0.00559313366221198
1.06934673366834 0.00477114702585545
1.07738693467337 0.0040552309520667
1.08542713567839 0.00343407525025081
1.09346733668342 0.0028972201374542
1.10150753768844 0.00243504241054911
1.10954773869347 0.00203873419216098
1.11758793969849 0.00170027529236702
1.12562814070352 0.00141240023434446
1.13366834170854 0.00116856097699961
1.14170854271357 0.000962886333322339
1.14974874371859 0.00079013903227832
1.15778894472362 0.000645671307204496
1.16582914572864 0.000525379817776158
1.17386934673367 0.00042566062856883
1.18190954773869 0.000343364877890817
1.18994974874372 0.000275755678621761
1.19798994974874 0.00022046670075852
1.20603015075377 0.000175462795476918
1.21407035175879 0.000139002934692111
1.22211055276382 0.000109605659932247
1.23015075376884 8.60171610708916e-05
1.23819095477387 6.71820399872046e-05
1.24623115577889 5.22167570974193e-05
1.25427135678392 4.03857101739794e-05
1.26231155778894 3.10798549072204e-05
1.27035175879397 2.37977449946597e-05
1.278391959799 1.81288456911768e-05
1.28643216080402 1.37389580898649e-05
1.29447236180905 1.0357581185868e-05
1.30251256281407 7.76703419045776e-06
1.3105527638191 5.79316176310571e-06
1.31859296482412 4.29744896866896e-06
1.32663316582915 3.17038002683164e-06
1.33467336683417 2.32588453461731e-06
1.3427135678392 1.69672611132734e-06
1.35075376884422 1.23070071972164e-06
1.35879396984925 8.87524725080542e-07
1.36683417085427 6.36305620885438e-07
1.3748743718593 4.53500918955941e-07
1.38291457286432 3.21282697814377e-07
1.39095477386935 2.26236525806323e-07
1.39899497487437 1.58333791739757e-07
1.4070351758794 1.10125809462358e-07
1.41507537688442 7.61163854555542e-08
1.42311557788945 5.22768599127029e-08
1.43115577889447 3.56739908399135e-08
1.4391959798995 2.41865075477685e-08
1.44723618090452 1.62907885912222e-08
1.45527638190955 1.09000018754517e-08
1.46331658291457 7.24426649218214e-09
1.4713567839196 4.78204135535599e-09
1.47939698492462 3.13509572762476e-09
1.48743718592965 2.04114652137727e-09
1.49547738693467 1.31962520381961e-09
1.5035175879397 8.4712410283708e-10
1.51155778894472 5.39921081902101e-10
1.51959798994975 3.41638618214145e-10
1.52763819095477 2.14596922466123e-10
1.5356783919598 1.33803089897319e-10
1.54371859296482 8.28058457119022e-11
1.55175879396985 5.08597188916375e-11
1.55979899497487 3.10006459027382e-11
1.5678391959799 1.8750674937207e-11
1.57587939698492 1.12532459607642e-11
1.58391959798995 6.70067651374509e-12
1.59195979899497 3.95826111549588e-12
1.6 2.31953008230669e-12
};
\addplot [green!50.0!black, dotted]
table {%
0 1
0.00804020100502513 0.999995584129114
0.0160804020100503 0.999982341687373
0.0241206030150754 0.999960288183234
0.0321608040201005 0.999929449449828
0.0402010050251256 0.999889861623525
0.0482412060301508 0.999841571113945
0.0562814070351759 0.999784634565427
0.064321608040201 0.999719118809998
0.0723618090452261 0.999645100811847
0.0804020100502513 0.999562667603367
0.0884422110552764 0.99947191621278
0.0964824120603015 0.999372953583407
0.104522613065327 0.999265896484627
0.112562814070352 0.999150871414579
0.120603015075377 0.999028014494669
0.128643216080402 0.99889747135594
0.136683417085427 0.998759397017387
0.144723618090452 0.998613955756263
0.152763819095477 0.998461320970483
0.160804020100503 0.998301675033176
0.168844221105528 0.998135209139487
0.176884422110553 0.997962123145702
0.184924623115578 0.997782625400793
0.192964824120603 0.997596932570461
0.201005025125628 0.99740526945378
0.209045226130653 0.997207868792533
0.217085427135678 0.997004971073325
0.225125628140704 0.996796824322588
0.233165829145729 0.99658368389456
0.241206030150754 0.996365812252334
0.249246231155779 0.996143478742091
0.257286432160804 0.995916959360599
0.265326633165829 0.995686536516075
0.273366834170854 0.995452498782518
0.281407035175879 0.995215140647594
0.289447236180905 0.994974762254172
0.29748743718593 0.994731669135605
0.305527638190955 0.994486171944835
0.31356783919598 0.994238586177416
0.321608040201005 0.993989231888535
0.32964824120603 0.993738433404106
0.337688442211055 0.993486519026021
0.34572864321608 0.99323382073162
0.353768844221106 0.992980673867451
0.361809045226131 0.992727416837385
0.369849246231156 0.992474390785134
0.377889447236181 0.992221939271231
0.385929648241206 0.991970407944519
0.393969849246231 0.991720144208172
0.402010050251256 0.991471496880309
0.410050251256281 0.9912248158492
0.418090452261307 0.990980451723098
0.426130653266332 0.990738755474709
0.434170854271357 0.990500078080301
0.442211055276382 0.990264770153445
0.450251256281407 0.990033181573395
0.458291457286432 0.989805661108063
0.466331658291457 0.989582556031585
0.474371859296482 0.989364211736429
0.482412060301508 0.989150971340007
0.490452261306533 0.988943175285734
0.498492462311558 0.988741160938486
0.506532663316583 0.988545262174366
0.514572864321608 0.988355808964717
0.522613065326633 0.988173126954291
0.530653266331658 0.987997537033471
0.538693467336683 0.987829354904446
0.546733668341709 0.987668890641232
0.554773869346734 0.987516448243395
0.562814070351759 0.987372325183376
0.570854271356784 0.987236811947252
0.578894472361809 0.987110191568793
0.586934673366834 0.986992739156678
0.594974874371859 0.986884721414669
0.603015075376884 0.986786396154614
0.61105527638191 0.986698011802074
0.619095477386935 0.986619806894396
0.62713567839196 0.986552009571045
0.635175879396985 0.986494837056003
0.64321608040201 0.986448495132021
0.651256281407035 0.986413177606527
0.65929648241206 0.986389065768979
0.667336683417086 0.986376327839447
0.675376884422111 0.986375118408208
0.683417085427136 0.986385577866144
0.691457286432161 0.986407831825705
0.699497487437186 0.986441990532243
0.707537688442211 0.986488148265478
0.715577889447236 0.986546382730894
0.723618090452261 0.986616754440839
0.731658291457287 0.986699306085135
0.739698492462312 0.986794061890987
0.747738693467337 0.986901026971989
0.755778894472362 0.987020186666057
0.763819095477387 0.987151505862089
0.771859296482412 0.987294928315208
0.779899497487437 0.987450375950415
0.787939698492462 0.987617748154536
0.795979899497487 0.98779692105632
0.804020100502513 0.987987746794602
0.812060301507538 0.988190052774448
0.820100502512563 0.98840364091122
0.828140703517588 0.988628286862537
0.836180904522613 0.988863739248125
0.844221105527638 0.989109718857582
0.852261306532663 0.989365917846134
0.860301507537689 0.98963199891846
0.868341708542714 0.989907594500741
0.876381909547739 0.990192305901108
0.884422110552764 0.990485702458712
0.892462311557789 0.990787320681699
0.900502512562814 0.991096663374419
0.908542713567839 0.991413198754247
0.916582914572864 0.991736359558469
0.92462311557789 0.992065542141746
0.932663316582915 0.992400105564726
0.94070351758794 0.992739370674464
0.948743718592965 0.99308261917738
0.95678391959799 0.993429092705553
0.964824120603015 0.993777991877277
0.97286432160804 0.99412847535283
0.980904522613065 0.994479658886574
0.988944723618091 0.994830614376561
0.996984924623116 0.995180368912934
1.00502512562814 0.995527903826536
1.01306532663317 0.99587215373925
1.02110552763819 0.996212005617712
1.02914572864322 0.99654629783218
1.03718592964824 0.996873819222459
1.04522613065327 0.997193308172944
1.05326633165829 0.997503451698964
1.06130653266332 0.997802884546791
1.06934673366834 0.998090188309799
1.07738693467337 0.99836389056347
1.08542713567839 0.998622464022051
1.09346733668342 0.998864325719913
1.10150753768844 0.999087836220776
1.10954773869347 0.99929129885819
1.11758793969849 0.999472959010856
1.12562814070352 0.999631003416527
1.13366834170854 0.999763559528486
1.14170854271357 0.999868694918746
1.14974874371859 0.999944416732381
1.15778894472362 0.999988671197561
1.16582914572864 0.999999343196113
1.17386934673367 0.999974255899636
1.18190954773869 0.999911170476413
1.18994974874372 0.999807785874606
1.19798994974874 0.999661738687404
1.20603015075377 0.999470603106062
1.21407035175879 0.999231890966954
1.22211055276382 0.998943051898988
1.23015075376884 0.998601473577966
1.23819095477387 0.998204482094655
1.24623115577889 0.997749342443551
1.25427135678392 0.997233259139516
1.26231155778894 0.996653376969659
1.27035175879397 0.996006781887989
1.278391959799 0.995290502060566
1.28643216080402 0.994501509068986
1.29447236180905 0.993636719280216
1.30251256281407 0.992692995390863
1.3105527638191 0.991667148154098
1.31859296482412 0.990555938297505
1.32663316582915 0.989356078640174
1.33467336683417 0.988064236417391
1.3427135678392 0.986677035821248
1.35075376884422 0.985191060765458
1.35879396984925 0.983602857882596
1.36683417085427 0.981908939761832
1.3748743718593 0.980105788435094
1.38291457286432 0.978189859119362
1.39095477386935 0.976157584222565
1.39899497487437 0.974005377620212
1.4070351758794 0.971729639209552
1.41507537688442 0.969326759747608
1.42311557788945 0.966793125978967
1.43115577889447 0.964125126058637
1.4391959798995 0.961319155274667
1.44723618090452 0.958371622074541
1.45527638190955 0.955278954398571
1.46331658291457 0.952037606322685
1.4713567839196 0.948644065012057
1.47939698492462 0.945094857986034
1.48743718592965 0.941386560693707
1.49547738693467 0.93751580439828
1.5035175879397 0.933479284367124
1.51155778894472 0.92927376836303
1.51959798994975 0.924896105430725
1.52763819095477 0.92034323497113
1.5356783919598 0.915612196094242
1.54371859296482 0.910700137239747
1.55175879396985 0.905604326052649
1.55979899497487 0.900322159499307
1.5678391959799 0.89485117420724
1.57587939698492 0.889189057010012
1.58391959798995 0.883333655676311
1.59195979899497 0.87728298980015
1.6 0.871035261826792
};
\end{axis}

\end{tikzpicture}

%% file: figures/darkfield/darkfield.tex

\definecolor{dGray}{gray}{0.7}
\definecolor{lGray}{gray}{0.9}
\definecolor{cb4b6b9}{RGB}{180,182,185}
\definecolor{c211e1f}{RGB}{33,30,31}
\definecolor{c77797b}{RGB}{119,121,123}
\def\y1{0.5}

\begin{tikzpicture}[y=0.80pt, x=0.80pt, yscale=-\y1, xscale=0.8, inner sep=0pt, outer sep=0pt]
	\begin{scope}[cm={{2.062,0.0,0.0,-2.062,(-129.99548,1482.0161)}}]

\path[fill=cb4b6b9,nonzero rule,color=lightgray] (283.7750,586.0280) .. controls
(283.7750,582.2590) and (248.4730,579.2040) .. (229.5980,579.2040) .. controls
(210.7230,579.2040) and (175.4200,582.2590) .. (175.4200,586.0280) .. controls
(175.4200,589.7960) and (210.7230,592.8500) .. (229.5980,592.8500) .. controls
(248.4730,592.8500) and (283.7750,589.7960) .. (283.7750,586.0280);
\path[draw=c211e1f,line join=miter,line cap=butt,miter limit=4.00,line
width=0.400pt] (283.7750,586.0280) .. controls (283.7750,582.2590) and
(248.4730,579.2040) .. (229.5980,579.2040) .. controls (210.7230,579.2040) and
(175.4200,582.2590) .. (175.4200,586.0280) .. controls (175.4200,589.7960) and
(210.7230,592.8500) .. (229.5980,592.8500) .. controls (248.4730,592.8500) and
(283.7750,589.7960) .. (283.7750,586.0280) -- cycle;

\path[fill=cb4b6b9,nonzero rule,pattern=north west lines,pattern color=lightgray] (283.0290,532.5260) .. controls
(283.0290,528.8980) and (248.7270,525.9580) .. (229.8510,525.9580) .. controls
(210.9750,525.9580) and (175.6740,528.8980) .. (175.6740,532.5260) .. controls
(175.6740,536.1530) and (210.9750,539.0930) .. (229.8510,539.0930) .. controls
(248.7270,539.0930) and (283.0290,536.1530) .. (283.0290,532.5260);
\path[draw=c211e1f,line join=miter,line cap=butt,miter limit=4.00,line
width=0.400pt] (283.0290,532.5260) .. controls (283.0290,528.8980) and
(248.7270,525.9580) .. (229.8510,525.9580) .. controls (210.9750,525.9580) and
(175.6740,528.8980) .. (175.6740,532.5260) .. controls (175.6740,536.1530) and
(210.9750,539.0930) .. (229.8510,539.0930) .. controls (248.7270,539.0930) and
(283.0290,536.1530) .. (283.0290,532.5260) -- cycle;

\path[fill=cb4b6b9,nonzero rule,pattern=north west lines,pattern color=lightgray] (283.9060,472.3160) .. controls
(283.9060,468.6220) and (248.6320,465.6260) .. (229.7890,465.6260) .. controls
(210.9470,465.6260) and (175.6720,468.6220) .. (175.6720,472.3160) .. controls
(175.6720,476.0110) and (210.9470,479.0060) .. (229.7890,479.0060) .. controls
(248.6320,479.0060) and (283.9060,476.0110) .. (283.9060,472.3160);
\path[draw=c211e1f,line join=miter,line cap=butt,miter limit=4.00,line
width=0.400pt] (283.9060,472.3160) .. controls (283.9060,468.6220) and
(248.6320,465.6260) .. (229.7890,465.6260) .. controls (210.9470,465.6260) and
(175.6720,468.6220) .. (175.6720,472.3160) .. controls (175.6720,476.0110) and
(210.9470,479.0060) .. (229.7890,479.0060) .. controls (248.6320,479.0060) and
(283.9060,476.0110) .. (283.9060,472.3160) -- cycle;

\path[fill=cb4b6b9,nonzero rule,pattern=north west lines,pattern color=lightgray] (283.9060,432.8440) .. controls
(283.9060,429.1500) and (248.6320,426.1550) .. (229.7890,426.1550) .. controls
(210.9470,426.1550) and (175.6720,429.1500) .. (175.6720,432.8440) .. controls
(175.6720,436.5390) and (210.9470,439.5350) .. (229.7890,439.5350) .. controls
(248.6320,439.5350) and (283.9060,436.5390) .. (283.9060,432.8440);
\path[draw=c211e1f,line join=miter,line cap=butt,miter limit=4.00,line
width=0.400pt] (283.9060,432.8440) .. controls (283.9060,429.1500) and
(248.6320,426.1550) .. (229.7890,426.1550) .. controls (210.9470,426.1550) and
(175.6720,429.1500) .. (175.6720,432.8440) .. controls (175.6720,436.5390) and
(210.9470,439.5350) .. (229.7890,439.5350) .. controls (248.6320,439.5350) and
(283.9060,436.5390) .. (283.9060,432.8440) -- cycle;

\path[fill=cb4b6b9,nonzero rule,color=lightgray] (283.9060,391.8650) .. controls
(283.9060,388.1700) and (248.6320,385.1750) .. (229.7890,385.1750) .. controls
(210.9470,385.1750) and (175.6720,388.1700) .. (175.6720,391.8650) .. controls
(175.6720,395.5590) and (210.9470,398.5550) .. (229.7890,398.5550) .. controls
(248.6320,398.5550) and (283.9060,395.5590) .. (283.9060,391.8650);

\node[color=red!80!black, anchor=base west, yshift=-8*\y1, xshift=5] at (286.9230,391.8650) {ADL};
\node[color=red!80!black, anchor=base west, yshift=6*\y1, xshift=5] at (286.9230,404.0820) {HP2};
\node[color=red!80!black, anchor=base west, yshift=8*\y1, xshift=5] at (286.9230,483.2910) {HP1};
\node[color=red!80!black, anchor=base west, yshift=-8*\y1, xshift=5] at (286.9230,432.8440) {TL22};
\node[color=red!80!black, anchor=base west, yshift=-8*\y1, xshift=5] at (286.9230,472.3160) {Tl21};
\node[color=red!80!black, anchor=base west, yshift=-8*\y1, xshift=5] at (286.9230,532.5260) {TL12};
\node[color=red!80!black, anchor=base west, yshift=-8*\y1, xshift=5] at (286.9230,586.0280) {TL11};

\node[color=red!80!black, anchor=base west, yshift=0, xshift=5] at (286.9230,322.0650) {SA plane};
\node[color=red!80!black, anchor=base west, yshift=0, xshift=5] at (286.9230,626.0810) {specimen};

\path[draw=c211e1f,line join=miter,line cap=butt,miter limit=4.00,line
width=0.400pt] (283.9060,391.8650) .. controls (283.9060,388.1700) and
(248.6320,385.1750) .. (229.7890,385.1750) .. controls (210.9470,385.1750) and
(175.6720,388.1700) .. (175.6720,391.8650) .. controls (175.6720,395.5590) and
(210.9470,398.5550) .. (229.7890,398.5550) .. controls (248.6320,398.5550) and
(283.9060,395.5590) .. (283.9060,391.8650) -- cycle;

\path[fill=c77797b,nonzero rule,pattern=north west lines,pattern color=lightgray] (175.7070,486.2910) -- (283.8730,486.2910) --
(283.8730,498.3750) -- (175.7070,498.3750) -- (175.7070,486.2910) -- cycle;
\path[draw=c211e1f,line join=miter,line cap=butt,miter limit=4.00,line
width=0.400pt] (283.8730,486.2910) -- (175.7070,486.2910) --
(175.7070,498.3750) -- (283.8730,498.3750) -- (283.8730,486.2910) -- cycle;
\path[fill=c77797b,nonzero rule,pattern=north west lines,pattern color=lightgray] (175.7070,406.0820) -- (283.8730,406.0820) --
(283.8730,418.1660) -- (175.7070,418.1660) -- (175.7070,406.0820) -- cycle;
\path[draw=c211e1f,line join=miter,line cap=butt,miter limit=4.00,line
width=0.400pt] (283.8730,406.0820) -- (175.7070,406.0820) --
(175.7070,418.1660) -- (283.8730,418.1660) -- (283.8730,406.0820) --
cycle(220.2560,587.3940) -- (215.4649,587.3940) -- (215.4649,602.3210) --
(220.2560,602.3210) -- (220.2560,587.3940) -- cycle(215.4650,602.3210) --
(220.2560,587.3940)
(215.5760,587.3940) --(220.2560,602.3210)
(243.8210,587.3940) -- (239.0309,587.3940) --
(239.0309,602.3210) -- (243.8210,602.3210) -- (243.8210,587.3940) --
cycle(239.0310,602.3210) -- (243.8210,587.3940)(239.1420,587.3940) --
(243.8210,602.3210)(217.5820,546.8440) -- (212.7910,546.8440) --
(212.7910,555.5315) -- (217.5820,555.5315) -- (217.5820,546.8440) --
cycle(212.7910,555.5310) -- (217.5820,546.8440)(212.9020,546.8440) --
(217.5820,555.5310)(246.3240,546.8440) -- (241.5340,546.8440) --
(241.5340,555.5315) -- (246.3240,555.5315) -- (246.3240,546.8440) --
cycle(241.5340,555.5310) -- (246.3240,546.8440)(241.6450,546.8440) --
(246.3240,555.5310)(217.5820,488.0800) -- (212.7910,488.0800) --
(212.7910,496.7670) -- (217.5820,496.7670) -- (217.5820,488.0800) --
cycle(212.7910,496.7670) -- (217.5820,488.0800)(212.9020,488.0800) --
(217.5820,496.7670)(246.3240,488.0800) -- (241.5340,488.0800) --
(241.5340,496.7670) -- (246.3240,496.7670) -- (246.3240,488.0800) --
cycle(241.5340,496.7670) -- (246.3240,488.0800)(241.6450,488.0800) --
(246.3240,496.7670)(217.5820,461.2320) -- (212.7910,461.2320) --
(212.7910,469.9190) -- (217.5820,469.9190) -- (217.5820,461.2320) --
cycle(212.7910,469.9190) -- (217.5820,461.2320)(212.9020,461.2320) --
(217.5820,469.9190)(246.3240,461.2320) -- (241.5340,461.2320) --
(241.5340,469.9190) -- (246.3240,469.9190) -- (246.3240,461.2320) --
cycle(241.5340,469.9190) -- (246.3240,461.2320)(241.6450,461.2320) --
(246.3240,469.9190)(217.5820,435.0540) -- (212.7910,435.0540) --
(212.7910,443.7420) -- (217.5820,443.7420) -- (217.5820,435.0540) --
cycle(212.7910,443.7420) -- (217.5820,435.0540)(212.9020,435.0540) --
(217.5820,443.7420)(246.3240,435.0540) -- (241.5340,435.0540) --
(241.5340,443.7420) -- (246.3240,443.7420) -- (246.3240,435.0540) --
cycle(241.5340,443.7420) -- (246.3240,435.0540)(241.6450,435.0540) --
(246.3240,443.7420)(217.5820,407.7510) -- (212.7910,407.7510) --
(212.7910,416.4390) -- (217.5820,416.4390) -- (217.5820,407.7510) --
cycle(212.7910,416.4390) -- (217.5820,407.7510)(212.9020,407.7510) --
(217.5820,416.4390)(246.3240,407.7510) -- (241.5340,407.7510) --
(241.5340,416.4390) -- (246.3240,416.4390) -- (246.3240,407.7510) --
cycle(241.5340,416.4390) -- (246.3240,407.7510)(241.6450,407.7510) --
(246.3240,416.4390)(217.5820,370.6620) -- (212.7910,370.6620) --
(212.7910,379.3490) -- (217.5820,379.3490) -- (217.5820,370.6620) --
cycle(212.7910,379.3490) -- (217.5820,370.6620)(212.9020,370.6620) --
(217.5820,379.3490)(246.3240,370.6620) -- (241.5340,370.6620) --
(241.5340,379.3490) -- (246.3240,379.3490) -- (246.3240,370.6620) --
cycle(241.5340,379.3490) -- (246.3240,370.6620)(241.6450,370.6620) --
(246.3240,379.3490)(217.5820,342.5840) -- (212.7910,342.5840) --
(212.7910,351.2720) -- (217.5820,351.2720) -- (217.5820,342.5840) --
cycle(212.7910,351.2720) -- (217.5820,342.5840)(212.9020,342.5840) --
(217.5820,351.2720)(246.3240,342.5840) -- (241.5340,342.5840) --
(241.5340,351.2720) -- (246.3240,351.2720) -- (246.3240,342.5840) --
cycle(241.5340,351.2720) -- (246.3240,342.5840)(241.6450,342.5840) --
(246.3240,351.2720)
(229.4670,626.0810) -- (229.4670,306.5820);

\node[color=red!80!black, anchor=base west, yshift=-12*\y1, xshift=-30] at (166.0970,379.3490) {ISh};
\node[color=red!80!black, anchor=base west, yshift=-12*\y1, xshift=-30] at (166.0970,416.4390) {DPH2};
\node[color=red!80!black, anchor=base west, yshift=-12*\y1, xshift=-30] at (166.0970,443.7420) {DP22};
\node[color=red!80!black, anchor=base west, yshift=-12*\y1, xshift=-30] at (166.0970,469.9190) {DP21};
\node[color=red!80!black, anchor=base west, yshift=-12*\y1, xshift=-30] at (166.0970,496.7670) {DPH1};
\node[color=red!80!black, anchor=base west, yshift=-12*\y1, xshift=-30] at (166.0970,555.5310) {DP12};

\node[color=red!80!black, anchor=base west, yshift=-3, xshift=-30] at (166.0970,338.8890) {DStig};
\node[color=red!80!black, anchor=base west, yshift=-3, xshift=-30] at (166.0970,355.7310) {DSh};
\node[color=red!80!black, anchor=base west, yshift=-3, xshift=-30] at (166.0970,582.0560) {DP11};
\node[color=red!80!black, anchor=base west, yshift=-3, xshift=-30] at (166.0970,594.7750) {QPol};
\node[color=red!80!black, anchor=base west, yshift=-3, xshift=-30] at (166.0970,609.0010) {HPol};

\path[fill=c211e1f,nonzero rule] (229.4670,303.5640) .. controls
(228.9400,304.9840) and (228.0410,306.7460) .. (227.0900,307.8370) --
(229.4670,306.9770) -- (231.8440,307.8370) .. controls (230.8930,306.7460) and
(229.9930,304.9840) .. (229.4670,303.5640);

\path[draw=c211e1f,line join=miter,line cap=butt,miter limit=4.00,line
width=0.400pt] 
(173.0060,322.0650)--(285.9290,322.0650)    
(173.0060,626.0810) -- (285.9290,626.0810)  

(165.9560,355.7310) -- (168.9040,355.7310) -- (168.9040,338.8890) -- (166.0970,338.8890)
(168.9040,347.1690) -- (171.9910,347.1690)
(165.9560,609.0010) -- (168.9040,609.0010) -- (168.9040,582.0560) -- (166.0970,582.0560)
(168.9040,594.7750) -- (171.9910,594.7750);



\path[draw=c211e1f,line join=miter,line cap=butt,miter limit=4.00,line
width=0.400pt, color=black,rounded corners=5pt]
(249.4670,626.0810) -- (249.4670,585.9140) -- (180,390) -- (229.4670,322.0650);

\path[draw=c211e1f,line join=miter,line cap=butt,miter limit=4.00,line
width=0.400pt, color=red!80!black,rounded corners=5pt]
(229.4670,626.0810)-- (255,586.0810) -- (235,390) -- (215,322.0650)
(249.4670,626.0810)-- (275,586.0810) -- (182,390) -- (215,322.0650);



\end{scope}
\end{tikzpicture}

%% file: figures/calibtemp/calibtemp_2.tex
\pgfmathsetlength{\imagewidth}{.49\columnwidth}
\pgfmathsetlength{\imagescale}{\imagewidth/1024}

      \begin{tikzpicture}[x=\imagescale,y=-\imagescale]
      \sffamily
        \node[anchor=north west,inner sep=0pt,outer sep=0pt] at (0,0)
        {\includegraphics[width=\imagewidth]{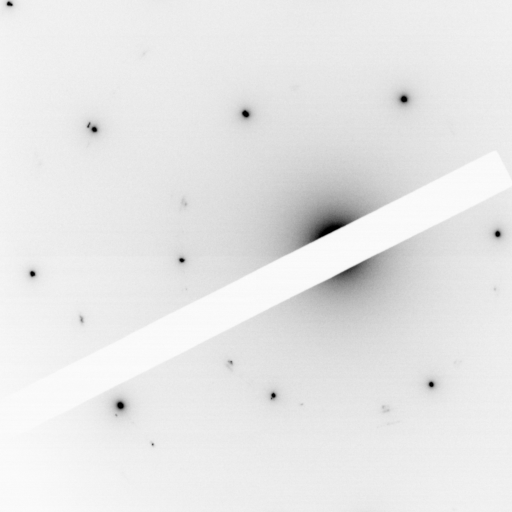}};
        \draw[ultra thick] (24,1000) node [anchor=south west] {2\:nm$^{\text{--1}}$} -- (250,1000);
        \node at (0,0) [anchor=north west] {(a)} ;
        \node at (1024,0) [anchor=north east] {130\:K} ;
      \end{tikzpicture}%
      \hfill
      \begin{tikzpicture}[x=\imagescale,y=-\imagescale]
      \sffamily
        \node[anchor=north west,inner sep=0pt,outer sep=0pt] at (0,0)
        {\includegraphics[width=\imagewidth]{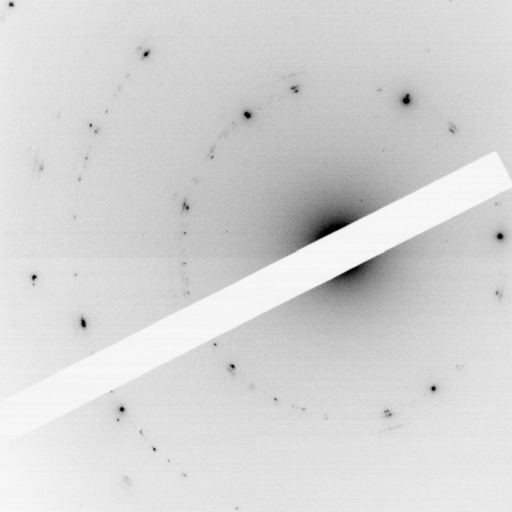}};
        \draw[] (768,303) node [anchor=south east] {$q_{\frac{2}{3}}$} circle (20);
        \draw[] (724,402) node [anchor=south east] {$q_{\frac{1}{3}}$} circle (20);
        \node at (0,0) [anchor=north west] {(b)} ;
        \node at (1024,0) [anchor=north east] {120\:K} ;
      \end{tikzpicture}%

%% file: manuscript.bbl
\begin{thebibliography}{29}%
\makeatletter
\providecommand \@ifxundefined [1]{%
 \@ifx{#1\undefined}
}%
\providecommand \@ifnum [1]{%
 \ifnum #1\expandafter \@firstoftwo
 \else \expandafter \@secondoftwo
 \fi
}%
\providecommand \@ifx [1]{%
 \ifx #1\expandafter \@firstoftwo
 \else \expandafter \@secondoftwo
 \fi
}%
\providecommand \natexlab [1]{#1}%
\providecommand \enquote  [1]{``#1''}%
\providecommand \bibnamefont  [1]{#1}%
\providecommand \bibfnamefont [1]{#1}%
\providecommand \citenamefont [1]{#1}%
\providecommand \href@noop [0]{\@secondoftwo}%
\providecommand \href [0]{\begingroup \@sanitize@url \@href}%
\providecommand \@href[1]{\@@startlink{#1}\@@href}%
\providecommand \@@href[1]{\endgroup#1\@@endlink}%
\providecommand \@sanitize@url [0]{\catcode `\\12\catcode `\$12\catcode
  `\&12\catcode `\#12\catcode `\^12\catcode `\_12\catcode `\%12\relax}%
\providecommand \@@startlink[1]{}%
\providecommand \@@endlink[0]{}%
\providecommand \url  [0]{\begingroup\@sanitize@url \@url }%
\providecommand \@url [1]{\endgroup\@href {#1}{\urlprefix }}%
\providecommand \urlprefix  [0]{URL }%
\providecommand \Eprint [0]{\href }%
\providecommand \doibase [0]{http://dx.doi.org/}%
\providecommand \selectlanguage [0]{\@gobble}%
\providecommand \bibinfo  [0]{\@secondoftwo}%
\providecommand \bibfield  [0]{\@secondoftwo}%
\providecommand \translation [1]{[#1]}%
\providecommand \BibitemOpen [0]{}%
\providecommand \bibitemStop [0]{}%
\providecommand \bibitemNoStop [0]{.\EOS\space}%
\providecommand \EOS [0]{\spacefactor3000\relax}%
\providecommand \BibitemShut  [1]{\csname bibitem#1\endcsname}%
\let\auto@bib@innerbib\@empty
\bibitem [{\citenamefont {Haider}\ \emph {et~al.}(1998)\citenamefont {Haider},
  \citenamefont {Uhlemann}, \citenamefont {Schwan}, \citenamefont {Rose},
  \citenamefont {Kabius},\ and\ \citenamefont {Urban}}]{Haider1998}%
  \BibitemOpen
  \bibfield  {author} {\bibinfo {author} {\bibfnamefont {M.}~\bibnamefont
  {Haider}}, \bibinfo {author} {\bibfnamefont {S.}~\bibnamefont {Uhlemann}},
  \bibinfo {author} {\bibfnamefont {E.}~\bibnamefont {Schwan}}, \bibinfo
  {author} {\bibfnamefont {H.}~\bibnamefont {Rose}}, \bibinfo {author}
  {\bibfnamefont {B.}~\bibnamefont {Kabius}}, \ and\ \bibinfo {author}
  {\bibfnamefont {K.}~\bibnamefont {Urban}},\ }\href@noop {} {\bibfield
  {journal} {\bibinfo  {journal} {Nature}\ }\textbf {\bibinfo {volume} {392}},\
  \bibinfo {pages} {768} (\bibinfo {year} {1998})}\BibitemShut {NoStop}%
\bibitem [{\citenamefont {Reimer}\ and\ \citenamefont
  {Kohl}(2008)}]{Reimer2008}%
  \BibitemOpen
  \bibfield  {author} {\bibinfo {author} {\bibfnamefont {L.}~\bibnamefont
  {Reimer}}\ and\ \bibinfo {author} {\bibfnamefont {H.}~\bibnamefont {Kohl}},\
  }\href@noop {} {\emph {\bibinfo {title} {Transmission Electron Microscopy:
  Physics of Image Formation}}}\ (\bibinfo  {publisher} {Springer Science \&
  Business Media},\ \bibinfo {year} {2008})\BibitemShut {NoStop}%
\bibitem [{\citenamefont {Bosman}\ \emph {et~al.}(2007)\citenamefont {Bosman},
  \citenamefont {Keast}, \citenamefont {Garc\'{\i}a-Mu\~noz}, \citenamefont
  {D'Alfonso}, \citenamefont {Findlay},\ and\ \citenamefont
  {Allen}}]{Bosman2007}%
  \BibitemOpen
  \bibfield  {author} {\bibinfo {author} {\bibfnamefont {M.}~\bibnamefont
  {Bosman}}, \bibinfo {author} {\bibfnamefont {V.~J.}\ \bibnamefont {Keast}},
  \bibinfo {author} {\bibfnamefont {J.~L.}\ \bibnamefont
  {Garc\'{\i}a-Mu\~noz}}, \bibinfo {author} {\bibfnamefont {A.~J.}\
  \bibnamefont {D'Alfonso}}, \bibinfo {author} {\bibfnamefont {S.~D.}\
  \bibnamefont {Findlay}}, \ and\ \bibinfo {author} {\bibfnamefont {L.~J.}\
  \bibnamefont {Allen}},\ }\href@noop {} {\bibfield  {journal} {\bibinfo
  {journal} {Physical Review Letters}\ }\textbf {\bibinfo {volume} {99}},\
  \bibinfo {pages} {086102} (\bibinfo {year} {2007})}\BibitemShut {NoStop}%
\bibitem [{\citenamefont {Egerton}(2009)}]{Egerton2009}%
  \BibitemOpen
  \bibfield  {author} {\bibinfo {author} {\bibfnamefont {R.~F.}\ \bibnamefont
  {Egerton}},\ }\href@noop {} {\bibfield  {journal} {\bibinfo  {journal}
  {Reports on Progress in Physics}\ }\textbf {\bibinfo {volume} {72}},\
  \bibinfo {pages} {016502} (\bibinfo {year} {2009})}\BibitemShut {NoStop}%
\bibitem [{\citenamefont {Frankel}\ and\ \citenamefont
  {Aitken}(1970)}]{Frankel1970}%
  \BibitemOpen
  \bibfield  {author} {\bibinfo {author} {\bibfnamefont {R.~S.}\ \bibnamefont
  {Frankel}}\ and\ \bibinfo {author} {\bibfnamefont {D.~W.}\ \bibnamefont
  {Aitken}},\ }\href@noop {} {\bibfield  {journal} {\bibinfo  {journal}
  {Applied Spectroscopy}\ }\textbf {\bibinfo {volume} {24}},\ \bibinfo {pages}
  {557} (\bibinfo {year} {1970})}\BibitemShut {NoStop}%
\bibitem [{\citenamefont {Lichte}(1991)}]{Lichte1991}%
  \BibitemOpen
  \bibfield  {author} {\bibinfo {author} {\bibfnamefont {H.}~\bibnamefont
  {Lichte}},\ }\href@noop {} {\bibfield  {journal} {\bibinfo  {journal}
  {Advances in Optical and Electron Microscopy}\ }\textbf {\bibinfo {volume}
  {12}},\ \bibinfo {pages} {25 } (\bibinfo {year} {1991})}\BibitemShut
  {NoStop}%
\bibitem [{\citenamefont {Linck}\ \emph {et~al.}(2012)\citenamefont {Linck},
  \citenamefont {Freitag}, \citenamefont {Kujawa}, \citenamefont {Lehmann},\
  and\ \citenamefont {Niermann}}]{Linck2012}%
  \BibitemOpen
  \bibfield  {author} {\bibinfo {author} {\bibfnamefont {M.}~\bibnamefont
  {Linck}}, \bibinfo {author} {\bibfnamefont {B.}~\bibnamefont {Freitag}},
  \bibinfo {author} {\bibfnamefont {S.}~\bibnamefont {Kujawa}}, \bibinfo
  {author} {\bibfnamefont {M.}~\bibnamefont {Lehmann}}, \ and\ \bibinfo
  {author} {\bibfnamefont {T.}~\bibnamefont {Niermann}},\ }\href@noop {}
  {\bibfield  {journal} {\bibinfo  {journal} {Ultramicroscopy}\ }\textbf
  {\bibinfo {volume} {116}},\ \bibinfo {pages} {13 } (\bibinfo {year}
  {2012})}\BibitemShut {NoStop}%
\bibitem [{\citenamefont {Winkler}\ \emph {et~al.}(2018)\citenamefont
  {Winkler}, \citenamefont {Barthel}, \citenamefont {Tavabi}, \citenamefont
  {Borghardt}, \citenamefont {Kardynal},\ and\ \citenamefont
  {Dunin-Borkowski}}]{Winkler2018}%
  \BibitemOpen
  \bibfield  {author} {\bibinfo {author} {\bibfnamefont {F.}~\bibnamefont
  {Winkler}}, \bibinfo {author} {\bibfnamefont {J.}~\bibnamefont {Barthel}},
  \bibinfo {author} {\bibfnamefont {A.~H.}\ \bibnamefont {Tavabi}}, \bibinfo
  {author} {\bibfnamefont {S.}~\bibnamefont {Borghardt}}, \bibinfo {author}
  {\bibfnamefont {B.~E.}\ \bibnamefont {Kardynal}}, \ and\ \bibinfo {author}
  {\bibfnamefont {R.~E.}\ \bibnamefont {Dunin-Borkowski}},\ }\href@noop {}
  {\bibfield  {journal} {\bibinfo  {journal} {Physical Review Letters}\
  }\textbf {\bibinfo {volume} {120}},\ \bibinfo {pages} {156101} (\bibinfo
  {year} {2018})}\BibitemShut {NoStop}%
\bibitem [{\citenamefont {Ashcroft}\ and\ \citenamefont
  {Mermin}(2011)}]{Ashcroft2011}%
  \BibitemOpen
  \bibfield  {author} {\bibinfo {author} {\bibfnamefont {N.~W.}\ \bibnamefont
  {Ashcroft}}\ and\ \bibinfo {author} {\bibfnamefont {N.~D.}\ \bibnamefont
  {Mermin}},\ }\href@noop {} {\emph {\bibinfo {title} {Solid State Physics}}}\
  (\bibinfo  {publisher} {Cengage Learning},\ \bibinfo {year}
  {2011})\BibitemShut {NoStop}%
\bibitem [{\citenamefont {Egerton}(2012)}]{Egerton2012}%
  \BibitemOpen
  \bibfield  {author} {\bibinfo {author} {\bibfnamefont {R.~F.}\ \bibnamefont
  {Egerton}},\ }\href@noop {} {\bibfield  {journal} {\bibinfo  {journal}
  {Microscopy Research and Technique}\ }\textbf {\bibinfo {volume} {75}},\
  \bibinfo {pages} {1550} (\bibinfo {year} {2012})}\BibitemShut {NoStop}%
\bibitem [{\citenamefont {Chiu}\ \emph {et~al.}(1986)\citenamefont {Chiu},
  \citenamefont {Downing}, \citenamefont {Dubochet}, \citenamefont {Glaeser},
  \citenamefont {Heide}, \citenamefont {Knapek}, \citenamefont {Kopf},
  \citenamefont {Lamvik}, \citenamefont {Lepault}, \citenamefont {Robertson},
  \citenamefont {Zeitler},\ and\ \citenamefont {Zemlin}}]{Chiu1986}%
  \BibitemOpen
  \bibfield  {author} {\bibinfo {author} {\bibfnamefont {W.}~\bibnamefont
  {Chiu}}, \bibinfo {author} {\bibfnamefont {K.~H.}\ \bibnamefont {Downing}},
  \bibinfo {author} {\bibfnamefont {J.}~\bibnamefont {Dubochet}}, \bibinfo
  {author} {\bibfnamefont {R.~M.}\ \bibnamefont {Glaeser}}, \bibinfo {author}
  {\bibfnamefont {H.~G.}\ \bibnamefont {Heide}}, \bibinfo {author}
  {\bibfnamefont {E.}~\bibnamefont {Knapek}}, \bibinfo {author} {\bibfnamefont
  {D.~A.}\ \bibnamefont {Kopf}}, \bibinfo {author} {\bibfnamefont {M.~K.}\
  \bibnamefont {Lamvik}}, \bibinfo {author} {\bibfnamefont {J.}~\bibnamefont
  {Lepault}}, \bibinfo {author} {\bibfnamefont {J.~D.}\ \bibnamefont
  {Robertson}}, \bibinfo {author} {\bibfnamefont {E.}~\bibnamefont {Zeitler}},
  \ and\ \bibinfo {author} {\bibfnamefont {F.}~\bibnamefont {Zemlin}},\
  }\href@noop {} {\bibfield  {journal} {\bibinfo  {journal} {Journal of
  Microscopy}\ }\textbf {\bibinfo {volume} {141}},\ \bibinfo {pages} {385}
  (\bibinfo {year} {1986})}\BibitemShut {NoStop}%
\bibitem [{\citenamefont {Laberrigue}\ and\ \citenamefont
  {Levinson}(1964)}]{Laberrigue1964}%
  \BibitemOpen
  \bibfield  {author} {\bibinfo {author} {\bibfnamefont {A.}~\bibnamefont
  {Laberrigue}}\ and\ \bibinfo {author} {\bibfnamefont {P.}~\bibnamefont
  {Levinson}},\ }\href@noop {} {\bibfield  {journal} {\bibinfo  {journal}
  {Comptes Rendus Hebdomadaires des Seances de l Academie des Sciences}\
  }\textbf {\bibinfo {volume} {259}},\ \bibinfo {pages} {530} (\bibinfo {year}
  {1964})}\BibitemShut {NoStop}%
\bibitem [{\citenamefont {Boersch}\ \emph {et~al.}(1966)\citenamefont
  {Boersch}, \citenamefont {Bostanjoglo},\ and\ \citenamefont
  {Lischke}}]{Boersch1966}%
  \BibitemOpen
  \bibfield  {author} {\bibinfo {author} {\bibfnamefont {H.}~\bibnamefont
  {Boersch}}, \bibinfo {author} {\bibfnamefont {O.}~\bibnamefont
  {Bostanjoglo}}, \ and\ \bibinfo {author} {\bibfnamefont {B.}~\bibnamefont
  {Lischke}},\ }\href@noop {} {\bibfield  {journal} {\bibinfo  {journal}
  {Optik}\ }\textbf {\bibinfo {volume} {24}},\ \bibinfo {pages} {460} (\bibinfo
  {year} {1966})}\BibitemShut {NoStop}%
\bibitem [{\citenamefont {{Gatan, Inc.}}(2018)}]{Gatan2018}%
  \BibitemOpen
  \bibfield  {author} {\bibinfo {author} {\bibnamefont {{Gatan, Inc.}}},\
  }\href
  {http://www.gatan.com/products/tem-specimen-holders/cooling-situ-holders}
  {\enquote {\bibinfo {title} {Cooling in-situ holders},}\ } (\bibinfo {year}
  {Date Accessed: Juli 05, 2018})\BibitemShut {NoStop}%
\bibitem [{\citenamefont {B\"orrnert}\ \emph {et~al.}(2015)\citenamefont
  {B\"orrnert}, \citenamefont {M\"uller}, \citenamefont {Riedel}, \citenamefont
  {Linck}, \citenamefont {Kirkland}, \citenamefont {Haider}, \citenamefont
  {B\"uchner},\ and\ \citenamefont {Lichte}}]{Boerrnert2015}%
  \BibitemOpen
  \bibfield  {author} {\bibinfo {author} {\bibfnamefont {F.}~\bibnamefont
  {B\"orrnert}}, \bibinfo {author} {\bibfnamefont {H.}~\bibnamefont
  {M\"uller}}, \bibinfo {author} {\bibfnamefont {T.}~\bibnamefont {Riedel}},
  \bibinfo {author} {\bibfnamefont {M.}~\bibnamefont {Linck}}, \bibinfo
  {author} {\bibfnamefont {A.~I.}\ \bibnamefont {Kirkland}}, \bibinfo {author}
  {\bibfnamefont {M.}~\bibnamefont {Haider}}, \bibinfo {author} {\bibfnamefont
  {B.}~\bibnamefont {B\"uchner}}, \ and\ \bibinfo {author} {\bibfnamefont
  {H.}~\bibnamefont {Lichte}},\ }\href@noop {} {\bibfield  {journal} {\bibinfo
  {journal} {Ultramicroscopy}\ }\textbf {\bibinfo {volume} {151}},\ \bibinfo
  {pages} {31 } (\bibinfo {year} {2015})}\BibitemShut {NoStop}%
\bibitem [{\citenamefont {B\"orrnert}\ \emph {et~al.}(2016)\citenamefont
  {B\"orrnert}, \citenamefont {Horst}, \citenamefont {Krzyzowski},\ and\
  \citenamefont {B\"uchner}}]{Boerrnert2016}%
  \BibitemOpen
  \bibfield  {author} {\bibinfo {author} {\bibfnamefont {F.}~\bibnamefont
  {B\"orrnert}}, \bibinfo {author} {\bibfnamefont {A.}~\bibnamefont {Horst}},
  \bibinfo {author} {\bibfnamefont {M.~A.}\ \bibnamefont {Krzyzowski}}, \ and\
  \bibinfo {author} {\bibfnamefont {B.}~\bibnamefont {B\"uchner}},\ }in\
  \href@noop {} {\emph {\bibinfo {booktitle} {European Microscopy Congress
  2016: Proceedings}}}\ (\bibinfo {year} {2016})\BibitemShut {NoStop}%
\bibitem [{\citenamefont {{B\"{o}rrnert}}\ \emph {et~al.}(2013)\citenamefont
  {{B\"{o}rrnert}}, \citenamefont {Bachmatiuk}, \citenamefont {Gorantla},
  \citenamefont {Wolf}, \citenamefont {Lubk}, \citenamefont {{B\"{u}chner}},\
  and\ \citenamefont {{R\"{u}mmeli}}}]{Boerrnert2013}%
  \BibitemOpen
  \bibfield  {author} {\bibinfo {author} {\bibfnamefont {F.}~\bibnamefont
  {{B\"{o}rrnert}}}, \bibinfo {author} {\bibfnamefont {A.}~\bibnamefont
  {Bachmatiuk}}, \bibinfo {author} {\bibfnamefont {S.}~\bibnamefont
  {Gorantla}}, \bibinfo {author} {\bibfnamefont {D.}~\bibnamefont {Wolf}},
  \bibinfo {author} {\bibfnamefont {A.}~\bibnamefont {Lubk}}, \bibinfo {author}
  {\bibfnamefont {B.}~\bibnamefont {{B\"{u}chner}}}, \ and\ \bibinfo {author}
  {\bibfnamefont {M.~H.}\ \bibnamefont {{R\"{u}mmeli}}},\ }\href@noop {}
  {\bibfield  {journal} {\bibinfo  {journal} {Journal of Microscopy}\ }\textbf
  {\bibinfo {volume} {249}},\ \bibinfo {pages} {87} (\bibinfo {year}
  {2013})}\BibitemShut {NoStop}%
\bibitem [{\citenamefont {Snoeck}\ \emph {et~al.}(2006)\citenamefont {Snoeck},
  \citenamefont {Hartel}, \citenamefont {M{\"u}ller},\ and\ \citenamefont
  {Tiemeijer}}]{Snoeck2006}%
  \BibitemOpen
  \bibfield  {author} {\bibinfo {author} {\bibfnamefont {E.}~\bibnamefont
  {Snoeck}}, \bibinfo {author} {\bibfnamefont {P.}~\bibnamefont {Hartel}},
  \bibinfo {author} {\bibfnamefont {H.}~\bibnamefont {M{\"u}ller}}, \ and\
  \bibinfo {author} {\bibfnamefont {P.~C.}\ \bibnamefont {Tiemeijer}},\ }in\
  \href@noop {} {\emph {\bibinfo {booktitle} {Proceedings of 16th International
  Microscopy Congress}}}\ (\bibinfo {year} {2006})\BibitemShut {NoStop}%
\bibitem [{\citenamefont {Houdellier}\ \emph {et~al.}(2008)\citenamefont
  {Houdellier}, \citenamefont {H{\"y}tch}, \citenamefont {H{\"u}e},\ and\
  \citenamefont {Snoeck}}]{Houdellier2008}%
  \BibitemOpen
  \bibfield  {author} {\bibinfo {author} {\bibfnamefont {F.}~\bibnamefont
  {Houdellier}}, \bibinfo {author} {\bibfnamefont {M.}~\bibnamefont
  {H{\"y}tch}}, \bibinfo {author} {\bibfnamefont {F.}~\bibnamefont {H{\"u}e}},
  \ and\ \bibinfo {author} {\bibfnamefont {E.}~\bibnamefont {Snoeck}},\
  }\href@noop {} {\bibfield  {journal} {\bibinfo  {journal} {Advances in
  Imaging and Electron Physics}\ }\textbf {\bibinfo {volume} {153}},\ \bibinfo
  {pages} {225} (\bibinfo {year} {2008})}\BibitemShut {NoStop}%
\bibitem [{\citenamefont {Haider}\ \emph {et~al.}(2009)\citenamefont {Haider},
  \citenamefont {Hartel}, \citenamefont {M{\"u}ller}, \citenamefont
  {Uhlemann},\ and\ \citenamefont {Zach}}]{Haider2009}%
  \BibitemOpen
  \bibfield  {author} {\bibinfo {author} {\bibfnamefont {M.}~\bibnamefont
  {Haider}}, \bibinfo {author} {\bibfnamefont {P.}~\bibnamefont {Hartel}},
  \bibinfo {author} {\bibfnamefont {H.}~\bibnamefont {M{\"u}ller}}, \bibinfo
  {author} {\bibfnamefont {S.}~\bibnamefont {Uhlemann}}, \ and\ \bibinfo
  {author} {\bibfnamefont {J.}~\bibnamefont {Zach}},\ }\href@noop {} {\bibfield
   {journal} {\bibinfo  {journal} {Philosophical Transactions of the Royal
  Society of London A}\ }\textbf {\bibinfo {volume} {367}},\ \bibinfo {pages}
  {3665} (\bibinfo {year} {2009})}\BibitemShut {NoStop}%
\bibitem [{\citenamefont {M{\"u}ller}\ \emph {et~al.}(2008)\citenamefont
  {M{\"u}ller}, \citenamefont {Uhlemann}, \citenamefont {Hartel},\ and\
  \citenamefont {Haider}}]{Mueller2008}%
  \BibitemOpen
  \bibfield  {author} {\bibinfo {author} {\bibfnamefont {H.}~\bibnamefont
  {M{\"u}ller}}, \bibinfo {author} {\bibfnamefont {S.}~\bibnamefont
  {Uhlemann}}, \bibinfo {author} {\bibfnamefont {P.}~\bibnamefont {Hartel}}, \
  and\ \bibinfo {author} {\bibfnamefont {M.}~\bibnamefont {Haider}},\
  }\href@noop {} {\bibfield  {journal} {\bibinfo  {journal} {Physics Procedia}\
  }\textbf {\bibinfo {volume} {1}},\ \bibinfo {pages} {167 } (\bibinfo {year}
  {2008})}\BibitemShut {NoStop}%
\bibitem [{\citenamefont {Uhlemann}\ \emph {et~al.}(2013)\citenamefont
  {Uhlemann}, \citenamefont {M\"uller}, \citenamefont {Hartel}, \citenamefont
  {Zach},\ and\ \citenamefont {Haider}}]{uhlemann2013}%
  \BibitemOpen
  \bibfield  {author} {\bibinfo {author} {\bibfnamefont {S.}~\bibnamefont
  {Uhlemann}}, \bibinfo {author} {\bibfnamefont {H.}~\bibnamefont {M\"uller}},
  \bibinfo {author} {\bibfnamefont {P.}~\bibnamefont {Hartel}}, \bibinfo
  {author} {\bibfnamefont {J.}~\bibnamefont {Zach}}, \ and\ \bibinfo {author}
  {\bibfnamefont {M.}~\bibnamefont {Haider}},\ }\href@noop {} {\bibfield
  {journal} {\bibinfo  {journal} {Physical Review Letters}\ }\textbf {\bibinfo
  {volume} {111}},\ \bibinfo {pages} {046101} (\bibinfo {year}
  {2013})}\BibitemShut {NoStop}%
\bibitem [{\citenamefont {{Kleindiek Nanotechnik
  GmbH}}(2017)}]{Kleindiek_2017}%
  \BibitemOpen
  \bibfield  {author} {\bibinfo {author} {\bibnamefont {{Kleindiek Nanotechnik
  GmbH}}},\ }\href {http://www.kleindiek.com/lt3310.html} {\enquote {\bibinfo
  {title} {Kleindiek nanotechnik: Substage with 10 mm travel},}\ } (\bibinfo
  {year} {Date Accessed: September 04, 2017})\BibitemShut {NoStop}%
\bibitem [{\citenamefont {Frank}(1973)}]{Frank1973}%
  \BibitemOpen
  \bibfield  {author} {\bibinfo {author} {\bibfnamefont {J.}~\bibnamefont
  {Frank}},\ }\href@noop {} {\bibfield  {journal} {\bibinfo  {journal} {Optik}\
  }\textbf {\bibinfo {volume} {38}},\ \bibinfo {pages} {519} (\bibinfo {year}
  {1973})}\BibitemShut {NoStop}%
\bibitem [{\citenamefont {Lichte}(1996)}]{Lichte1996}%
  \BibitemOpen
  \bibfield  {author} {\bibinfo {author} {\bibfnamefont {H.}~\bibnamefont
  {Lichte}},\ }\href@noop {} {\bibfield  {journal} {\bibinfo  {journal}
  {Ultramicroscopy}\ }\textbf {\bibinfo {volume} {64}},\ \bibinfo {pages} {79 }
  (\bibinfo {year} {1996})}\BibitemShut {NoStop}%
\bibitem [{\citenamefont {Sickmann}\ \emph {et~al.}(2011)\citenamefont
  {Sickmann}, \citenamefont {Form{\'a}nek}, \citenamefont {Linck},
  \citenamefont {Muehle},\ and\ \citenamefont {Lichte}}]{Sickmann2011}%
  \BibitemOpen
  \bibfield  {author} {\bibinfo {author} {\bibfnamefont {J.}~\bibnamefont
  {Sickmann}}, \bibinfo {author} {\bibfnamefont {P.}~\bibnamefont
  {Form{\'a}nek}}, \bibinfo {author} {\bibfnamefont {M.}~\bibnamefont {Linck}},
  \bibinfo {author} {\bibfnamefont {U.}~\bibnamefont {Muehle}}, \ and\ \bibinfo
  {author} {\bibfnamefont {H.}~\bibnamefont {Lichte}},\ }\href@noop {}
  {\bibfield  {journal} {\bibinfo  {journal} {Ultramicroscopy}\ }\textbf
  {\bibinfo {volume} {111}},\ \bibinfo {pages} {290 } (\bibinfo {year}
  {2011})}\BibitemShut {NoStop}%
\bibitem [{\citenamefont {Lehmann}(2004)}]{Lehmann2004}%
  \BibitemOpen
  \bibfield  {author} {\bibinfo {author} {\bibfnamefont {M.}~\bibnamefont
  {Lehmann}},\ }\href@noop {} {\bibfield  {journal} {\bibinfo  {journal}
  {Ultramicroscopy}\ }\textbf {\bibinfo {volume} {100}},\ \bibinfo {pages} {9 }
  (\bibinfo {year} {2004})}\BibitemShut {NoStop}%
\bibitem [{\citenamefont {K{\"o}nig}(2013)}]{Koenig2013}%
  \BibitemOpen
  \bibfield  {author} {\bibinfo {author} {\bibfnamefont {A.}~\bibnamefont
  {K{\"o}nig}},\ }\emph {\bibinfo {title} {Charge-Density Waves and Collective
  Dynamics in the Transition-Metal Dichalcogenides: An Electron Energy-Loss
  Study}},\ \href@noop {} {Ph.D. thesis},\ \bibinfo  {school} {TU Dresden}
  (\bibinfo {year} {2013})\BibitemShut {NoStop}%
\bibitem [{\citenamefont {Moncton}\ \emph {et~al.}(1975)\citenamefont
  {Moncton}, \citenamefont {Axe},\ and\ \citenamefont {DiSalvo}}]{Moncton1975}%
  \BibitemOpen
  \bibfield  {author} {\bibinfo {author} {\bibfnamefont {D.~E.}\ \bibnamefont
  {Moncton}}, \bibinfo {author} {\bibfnamefont {J.~D.}\ \bibnamefont {Axe}}, \
  and\ \bibinfo {author} {\bibfnamefont {F.~J.}\ \bibnamefont {DiSalvo}},\
  }\href@noop {} {\bibfield  {journal} {\bibinfo  {journal} {Physical Review
  Letters}\ }\textbf {\bibinfo {volume} {34}},\ \bibinfo {pages} {734}
  (\bibinfo {year} {1975})}\BibitemShut {NoStop}%
\end{thebibliography}
